\documentclass[camera,letterpaper,nomarginnotes,nonarrowgutter]{jpaper}
\usepackage{xspace}
\usepackage{tikz}
\usepackage{xcolor}
\usepackage{soul}
\usepackage{multicol}
\usepackage{multirow}
\usepackage{booktabs}
\usepackage{fancyhdr}
\usepackage{pdfpages}
\usepackage[nomessages]{fp}
\usepackage{tocbibind}
\usepackage{clipboard}

\usepackage[noadjust]{cite}
\usepackage{amsmath,amssymb,amsfonts}
\usepackage{algorithmic}
\usepackage{graphicx}
\usepackage{textcomp}
\usepackage{balance}
\usepackage{mathptmx} 
\usepackage[normalem]{ulem}
\usepackage{url}
\usepackage{marginnote}
\usepackage{enumitem}
\usepackage{ifthen}
\usepackage{caption}
\usepackage{listings}
\usepackage{setspace}
\usepackage{float}
\usepackage[us,12hr]{datetime}
\usepackage{xargs}
\usepackage{titletoc}
\usepackage[all]{nowidow}
\usepackage[framemethod=tikz]{mdframed}
\usepackage[most]{tcolorbox}
\tcbuselibrary{breakable}
\tcbuselibrary{hooks}%YOU MUST USE THIS FOR <extras last pre> KEY
\usepackage{mathtools}

\usepackage{courier}
\usepackage[italic]{mathastext}

\usepackage{siunitx}
\usepackage{algorithm}
\usepackage{arydshln} % for dashed line in table

\usepackage{pifont}
\usepackage{footmisc}
\usepackage{hhline} % for table double lines that don't interrupt vertical lines
\usepackage[compact]{titlesec} % squeeze title spacings
\usepackage{footnote}

\usepackage{circledsteps}

\usepackage[bookmarks=true,breaklinks=true,colorlinks,linkcolor=black,citecolor=blue,urlcolor=black]{hyperref}
\usepackage[resetlabels]{multibib}

\definecolor{denim}{rgb}{0.08, 0.38, 0.74}
\definecolor{darkolivegreen}{rgb}{0.33, 0.42, 0.18}
\definecolor{dgreen}{rgb}{0.00, 0.75, 0.00}
\definecolor{darkpink}{rgb}{0.88, 0.28, 0.54}
\definecolor{forestgreen}{rgb}{0.0, 0.27, 0.13}
\definecolor{amber}{rgb}{1.0, 0.49, 0.0}
\definecolor{lightyellow}{rgb}{0.980, 0.956, 0.623}
\definecolor{lightblue}{rgb}{0.980, 0.956, 0.623}
\definecolor{darkamber}{rgb}{0.5, 0.19, 0.0}
\definecolor{dkgreen}{rgb}{0,0.6,0}
\definecolor{gray}{rgb}{0.5,0.5,0.5}
\definecolor{mauve}{rgb}{0.58,0,0.82}
\definecolor{lightmauve}{rgb}{0.68,0.4,0.92}
\definecolor{chocolate}{rgb}{0.48, 0.25, 0.0}
\definecolor{dollarbill}{rgb}{0.52,0.73,0.4}
\definecolor{dkdkgreen}{rgb}{0,0.45,0}
\definecolor{gfored}{rgb}{0.580, 0.050, 0.211}
\definecolor{darkwarmgray}{rgb}{0.15, 0.050, 0.05}
\definecolor{ups-truck}{rgb}{0.53, 0.28, 0.21}

\newcommand\rev[1]{{\color{black}{#1}}}

\newcommand\mech{RawHash\xspace}
\newcommand\mechcap{RawHash\xspace}

\newcommand\ltitle{RawHash: Enabling Fast and Accurate Real-Time Analysis\\ of Raw Nanopore Signals for Large Genomes\xspace}

\newcommand{\release}{\href{https://github.com/CMU-SAFARI/RawHash}{https://github.com/CMU-SAFARI/RawHash}\xspace}

\newcommand\unc{UNCALLED\xspace}
\newcommand\sig{Sigmap\xspace}
\newcommand\ru{Read Until\xspace}

\newcommand\avgthrU{$25.8\times$\xspace}

\newcommand\avgthrS{$3.4\times$\xspace}

\newcommand\avgmtU{$32.1\times$\xspace}

\newcommand\avgmtS{$2.1\times$\xspace}

\newcommand{\head}[1]{{\noindent\textbf{#1.}\xspace}} % for heading of a paragraph
\let\oldmarginnote\marginnote

\renewcommand{\marginnote}[2][rectangle,draw,fill=blue!40,rounded corners]{%
    \oldmarginnote{%
    \tikz \node at (0,0) [#1]{#2};}%
}

\titlespacing*{\section}{0pt}{0ex}{0ex}
\titlespacing*{\subsection}{0pt}{0ex}{0ex}
\titlespacing*{\subsubsection}{0pt}{0ex}{0ex}
\titlespacing\section{1pt}{5pt plus 0.5pt minus 4pt}{5pt plus 0.5pt minus 4pt}
\titlespacing\subsection{1pt}{5pt plus 0.5pt minus 3pt}{5pt plus 0.5pt minus 3pt}
\titlespacing\subsubsection{1pt}{5pt plus 0.5pt minus 2pt}{5pt plus 0.5pt minus 2pt}

\makeatletter
\g@addto@macro{\normalsize}{%
  \setlength{\abovedisplayskip}{2pt plus 1pt minus 1pt}
  \setlength{\belowdisplayskip}{2pt plus 1pt minus 1pt}
  \setlength{\abovedisplayshortskip}{0pt}
  \setlength{\belowdisplayshortskip}{0pt}
  \setlength{\intextsep}{2pt plus 1pt minus 1pt}
  \setlength{\textfloatsep}{3pt plus 1pt minus 1pt}
  \setlength{\dbltextfloatsep}{3pt plus 1pt minus 1pt}
  \setlength{\skip\footins}{4pt plus 1pt minus 1pt}}
\makeatother

\expandafter\def\expandafter\UrlBreaks\expandafter{\UrlBreaks
  \do\a\do\b\do\c\do\d\do\e\do\f\do\g\do\h\do\i\do\j
  \do\k\do\l\do\m\do\n\do\o\do\p\do\q\do\r\do\s\do\t
  \do\u\do\v\do\w\do\x\do\y\do\z\do\A\do\B\do\C\do\D
  \do\E\do\F\do\G\do\H\do\I\do\J\do\K\do\L\do\M\do\N
  \do\O\do\P\do\Q\do\R\do\S\do\T\do\U\do\V\do\W\do\X
  \do\Y\do\Z}

\newcommand{\squishlist}{
 \begin{list}{$\circ$}
  { \setlength{\itemsep}{0pt}
     \setlength{\parsep}{0pt}
     \setlength{\topsep}{0pt}
     \setlength{\partopsep}{0pt}
     \setlength{\leftmargin}{1em}
     \setlength{\labelwidth}{1em}
     \setlength{\labelsep}{0.5em} } }

\newcommand{\squishsublist}{
\begin{list}{$\rightarrow$}
 { \setlength{\itemsep}{0pt}
    \setlength{\parsep}{0pt}
    \setlength{\topsep}{-10em}
    \setlength{\partopsep}{-3pt}
    \setlength{\leftmargin}{1em}
    \setlength{\labelwidth}{1em}
    \setlength{\labelsep}{0.5em} } }

\newcommand{\squishend}{\end{list}}
  
\let\oldmarginnote\marginnote
\renewcommand{\marginnote}[2][rectangle,draw,fill=blue!40,rounded corners]{%
    \oldmarginnote{%
    \tikz \node at (0,0) [#1]{#2};}%
}

\mdfdefinestyle{graybox}{
    splittopskip=0,%
    splitbottomskip=0,%
    frametitleaboveskip=0,
    frametitlebelowskip=0,
    skipabove=0,%
    skipbelow=0,%
    leftmargin=0,%
    rightmargin=0,%
    innertopmargin=0mm,%
    innerbottommargin=0mm,%
    roundcorner=1mm,%
    backgroundcolor=lightblue,
    hidealllines=true}
    
\mdfdefinestyle{graybox2}{
    splittopskip=0,%
    splitbottomskip=0,%
    frametitleaboveskip=0,
    frametitlebelowskip=0,
    skipabove=0,%
    skipbelow=0,%
    leftmargin=0,%
    rightmargin=0,%
    innertopmargin=2mm,%
    innerbottommargin=2mm,%
    roundcorner=2mm,%
    backgroundcolor=lightblue,
    hidealllines=true}

\newcommandx{\changev}[2][1=]{\todo[ linecolor=blue,backgroundcolor=blue!25,bordercolor=blue,#1,size=\scriptsize]{#2}}

\let\oldmarginnote\marginnote
\renewcommand{\marginnote}[2][rectangle,draw,fill=blue!40,rounded corners]{%
        \oldmarginnote{%
        \tikz \node at (0,0) [#1]{#2};}%
        }
        
\newcommand{\boxbegin} {
	\begin{tcolorbox}[enhanced, frame hidden, colback=gray!50, breakable]
}

\newcommand{\boxend} {
	\end{tcolorbox}
}

\newcommand{\yboxbegin} {
	\begin{tcolorbox}[breakable, enhanced, frame hidden, colback=yellow!50]
}

\newcommand{\yboxend} {
	\end{tcolorbox}
}

\mdfdefinestyle{graybox}{
    splittopskip=0,%
    splitbottomskip=0,%
    frametitleaboveskip=0,
    frametitlebelowskip=0,
    skipabove=0,%
    skipbelow=0,%
    leftmargin=0,%
    rightmargin=0,%
    innertopmargin=0mm,%
    innerbottommargin=0mm,%
    roundcorner=1mm,%
    backgroundcolor=lightblue,
    hidealllines=true}
    
\mdfdefinestyle{graybox2} {
    splittopskip=0,%
    splitbottomskip=0,%
    frametitleaboveskip=0,
    frametitlebelowskip=0,
    skipabove=0,%
    skipbelow=0,%
    leftmargin=0,%
    rightmargin=0,%
    innertopmargin=2mm,%
    innerbottommargin=2mm,%
    roundcorner=2mm,%
    backgroundcolor=lightblue,
    hidealllines=true}

\newcommand{\bboxbegin}{
    \begin{mdframed}[style=graybox]
}

\newcommand{\bboxend}{
    \end{mdframed}
}

\newcommand{\yyboxbegin}{
    \begin{mdframed}[style=graybox2]
}

\newcommand{\yyboxend}{
    \end{mdframed}
}

\let\oldtableofcontents\tableofcontents%remember the definition
\renewcommand\tableofcontents{
  \oldtableofcontents%use the standard toc
  \clearpage
}

\makeindex
\definecolor{dgreen}{rgb}{0.0, 0.70, 0.25}
\newlength{\bibitemsep}
\newlength{\bibparskip}
\let\oldthebibliography\thebibliography
\renewcommand\thebibliography[1]{%
  \oldthebibliography{#1}%
  \setlength{\parskip}{\bibitemsep}%
  \setlength{\itemsep}{\bibparskip}%
}

\usepackage{titlesec}
\titlespacing*{\section}{0pt}{0.5ex}{0.5ex} % Adjust spacing for sections
\titlespacing*{\subsection}{0pt}{0.4ex}{0.5ex} % Adjust spacing for subsections
\titlespacing*{\subsubsection}{0pt}{0.3ex}{0.3ex} % Adjust spacing for subsections

\newcommand{\affilETH}[0]{\small {$^1$}}
\title{\ltitle} 
\author{\vspace{-17pt}\\%
\fontsize{11}{12}\selectfont%
{Can Firtina\affilETH{}}\quad%
{Nika Mansouri Ghiasi\affilETH{}}\quad%
{Joel Lindegger\affilETH{}}\quad%
{Gagandeep Singh\affilETH{}}\quad%
\vspace{-1pt}\\%
\fontsize{11}{12}\selectfont%
{Meryem Banu Cavlak\affilETH{}}\quad
{Haiyu Mao\affilETH{}}\quad%
{Onur Mutlu\affilETH{}}\quad%
\vspace{-1pt}\\%
{\fontsize{10}{11}\selectfont
\affilETH\emph{ETH Zurich}%
}
\vspace{-16pt}\vspace{0.3em}}
\newcites{supp}{Supplementary References}
\newcites{rev}{Revision References}

\begin{document}
% \bstctlcite{IEEEexample:BSTcontrol}
\maketitle
\thispagestyle{plain}
\pagestyle{plain}
\setstretch{0.906}

%%%%%% -- PAPER CONTENT STARTS-- %%%%%%%%

\begin{abstract}
Nanopore sequencers generate electrical raw signals in real-time while sequencing long genomic strands.~These raw signals can be analyzed as they are generated, providing an opportunity for real-time genome analysis. An important feature of nanopore sequencing, \ru, can eject strands from sequencers without fully sequencing them, which provides opportunities to computationally reduce the sequencing time and cost. However, existing works utilizing \ru either 1)~require powerful computational resources that may not be available for portable sequencers or 2)~lack scalability for large genomes, rendering them inaccurate or ineffective.

We propose \mech, the first mechanism that can accurately and efficiently perform real-time analysis of nanopore raw signals for large genomes using a hash-based similarity search. To enable this, \mechcap ensures the signals corresponding to the same DNA content lead to the same hash value, regardless of the slight variations in these signals. \mechcap achieves an accurate hash-based similarity search via an effective quantization of the raw signals such that signals corresponding to the same DNA content have the same quantized value and, subsequently, the same hash value.

We evaluate \mech on three applications: 1)~read mapping, 2)~relative abundance estimation, and 3)~contamination analysis. Our evaluations show that \mech is the only tool that can provide high accuracy and high throughput for analyzing large genomes in real-time. When compared to the state-of-the-art techniques, \unc and \sig, \mech provides 1)~\avgthrU and \avgthrS better average throughput and 2)~significantly better accuracy for large genomes, respectively. Source code is available at \release.
\end{abstract}
\section{Introduction} \label{sec:introduction}

High-throughput sequencing (HTS) devices can generate a large amount of genomic data at a relatively low cost. HTS can be used to analyze a wide range of samples, from small amounts of DNA or RNA to entire genomes. Oxford Nanopore Technologies (ONT) is one of the most widely-used HTS technologies that can sequence long genomic regions, called \emph{reads}, with up to a few million bases. ONT devices use the nanopore sequencing technique, which involves passing a single DNA or RNA strand through a tiny pore, \emph{nanopore} or channel, at an average speed of 450 bases per second~\cite{kovaka_targeted_2021} and measuring the electrical current as the strand passes through. Nanopore sequencing enables two key features. First, nanopores provide the electrical raw signals in \emph{real-time} as the DNA strand passes through a nanopore. Second, nanopore sequencing provides a functionality, known as \emph{\ru}~\cite{loose_real-time_2016}, that can partially sequence DNA strands without fully sequencing them. These two features of nanopores provide opportunities for 1)~real-time genome analysis and 2)~significantly reducing sequencing time and cost.

Real-time analysis of nanopore raw signals using \ru can reduce the sequencing time and cost per read by terminating the sequencing whenever sequencing the full read is not necessary. The freed-up nanopore can then be used to sequence a different read. A purely computational mechanism can send a signal to \emph{eject} a read from a nanopore by reversing the voltage if the partial sequencing of a read meets certain conditions for particular genome analysis, such as 1)~reaching a desired coverage for a species in a sample~\cite{payne_readfish_2021} or 2)~identifying that a read does not originate from a certain genome of interest (i.e., a target region)~\cite{kovaka_targeted_2021, zhang_real-time_2021} and hence, does not need to be fully sequenced. By terminating the sequencing of reads that do not correspond to the target region, the sequencer can spend time and resources on higher coverage sequencing of the reads that correspond to the target. This process is referred to as \emph{nanopore adaptive sampling}. By providing high coverage at target regions and avoiding unessential sequencing of reads outside those regions, this approach can improve the quality of sequencing and the downstream analysis utilizing the obtained data.

To effectively utilize adaptive sampling in nanopore sequencing, it is crucial to have computational methods that can accurately analyze the raw output signals from nanopores in real-time. These methods must provide 1)~low latency and 2)~throughput matching or exceeding that of the sequencer~\cite{dunn2021squigglefilter, kovaka_targeted_2021, zhang_real-time_2021}. Several works propose adaptive sampling methods for real-time analysis of raw nanopore signals~\cite{edwards_real-time_2019, dunn2021squigglefilter, bao_squigglenet_2021, kovaka_targeted_2021, zhang_real-time_2021, payne_readfish_2021, shih_efficient_2022, ulrich_readbouncer_2022, senanayake_deepselectnet_2023, sadasivan_rapid_2023}. However, these works have three key limitations. First, most techniques mainly use powerful computational resources, such as GPUs~\cite{payne_readfish_2021, bao_squigglenet_2021}, or specialized hardware~\cite{dunn2021squigglefilter, shih_efficient_2022} due to the use of computationally-intensive algorithms such \rev{as basecalling as we explain in detail in Section~\ref{sec:relatedwork}}. This can make real-time genome analysis challenging for portable and low-cost nanopore-based sequencers, such as the ONT Flongle or MinION, which are not typically equipped with such resources. Therefore these techniques introduce challenges for using them in resource-constrained environments. Second, the sheer size of genomic data at the scale of large genomes (e.g., human genome) makes it challenging to process the data in real-time. This is because such large genomes require efficient and accurate similarity identification across a large number of regions. This renders many current methods~\cite{kovaka_targeted_2021, zhang_real-time_2021} inaccurate or useless for large genomes as they cannot either provide accurate results or match the throughput of nanopores for these genomes. Third, machine learning models used in past works~\cite{edwards_real-time_2019, payne_readfish_2021, bao_squigglenet_2021, ulrich_readbouncer_2022, senanayake_deepselectnet_2023} to analyze raw nanopore signals often require retraining or reconfiguring the model to improve accuracy for a certain experiment, which can be a barrier to flexibly and easily performing real-time analysis without retraining or reconfiguring these models. To our knowledge, there is no work that can efficiently and accurately perform real-time analysis of raw nanopore signals on a large scale (e.g., whole-genome analysis for human) without requiring powerful computational resources, which can easily and flexibly be applied to a wide range of applications that could benefit from real-time nanopore raw signal analysis.

Our \textbf{goal} is to enable efficient and accurate real-time genome analysis for large genomes. To this end, we propose \textbf{\emph{\mech}}, the \emph{first} mechanism that can efficiently and accurately perform real-time analysis of raw nanopore signals for large genomes in resource-contained environments. Unlike all the past works, \mech is the only mechanism that can efficiently scale to large genomes and perform accurate real-time genomic analysis without requiring computationally-intensive algorithms such as basecalling. Our \textbf{key idea} is to encode regions of the raw nanopore signal into hash values such that similar signal regions can efficiently be identified by matching their hash values, facilitating efficient similarity identification between signals. However, enabling accurate hashing-based similarity identification in the raw signal domain is challenging because raw signals corresponding to the same DNA content are unlikely to have exactly the same signal amplitudes. This is because the raw signals generated by nanopores can vary each time the same DNA fragment is sequenced due to several factors impacting nanopores during sequencing, such as variations in the properties of the nanopores or the conditions in which the sequencing is performed~\cite{david_nanocall_2017}. Although the similarity identification of raw signals is possible via calculating the Euclidean distance between a sequence of signals in a multi-dimensional space~\cite{zhang_real-time_2021}, such an approach can become impractical when dealing with larger sequences as the number of dimensions increases with the length of the sequences. This increase in dimensionality can lead to computational complexity and the curse of dimensionality, making it expensive and impractical.

To address these challenges, \mech provides three \textbf{key mechanisms} for efficient signal encoding and similarity identification. First, \mech encodes signal values that have a wider range of values into a smaller set of values using a quantization technique, such that signal values within a certain range are assigned to the same encoded value. This helps to alleviate the probability of having varying signal values for the same DNA content and enables \mech to directly match these values using a hashing technique. Second, \mech concatenates the quantized values of \emph{multiple} consecutive signals and generates a single hash value for them. The hashing mechanism enables \mech to efficiently identify similar signal regions of these consecutive signal values by directly matching their corresponding hash values. Representing many consecutive signals with a single hash value increases the size of the regions examined during similarity identification without suffering from the curse of dimensionality. Using larger regions can substantially reduce the number of possible matching regions that need to be examined. \mech is the \emph{first} work that can accurately use hash values in the raw signal domain, which enables using efficient data structures commonly-used used in the sequence domain (e.g., hash tables in minimap2~\cite{li_minimap2_2018}). Third, \mech uses an existing algorithm, known as chaining~\cite{li_minimap2_2018}, to find the colinear matches of hash values between signals to identify similar signal regions. These efficient and accurate mechanisms enable \mech to perform real-time genome analysis for large genomes.

While our proposed three key mechanisms have the potential to be used for various purposes in raw signal similarity identification, we design \mech as a tool for mapping nanopore raw signals to their corresponding reference genomes in real-time. \mechcap operates the mapping in two steps 1)~\emph{indexing} and 2)~\emph{mapping}. First, in the indexing step, \mech 1)~converts the reference genome sequence into \emph{expected} signal values by simulating the expected behavior of nanopores based on a previously-known model, 2)~generates the hash values from these signals, and 3)~stores the hash values in a hash table for efficient matching. Second, in the mapping step, \mech 1)~generates the hash values from the raw signals in a streaming fashion, 2)~queries the hash table from the indexing step with these hash values to find the matching regions in the reference genome with the same hash value, and 3)~performs chaining to find the similar region between the reference genome and the raw signal of a read.

\mechcap can utilize the unique functionalities of nanopore sequencing to reduce the sequencing time and cost in two ways. First, to avoid redundant sequencing and processing of each read, \mech can use \ru to eject a read before it is fully sequenced if \mech identifies that the sequenced portion of the read can already be mapped to a reference genome. Second, to perform a cost- and time-efficient relative abundance estimation, \mech can utilize \emph{Run Until} to \emph{fully} stop the \emph{entire} sequencing of all subsequent reads after sequencing a certain amount of reads that is sufficient to make an accurate relative abundance estimation. We refer to such usage during abundance estimation as \textbf{\emph{Sequence Until}}. Avoiding the redundant sequencing of further reads that are unlikely to substantially change the relative abundance estimation has the potential to significantly reduce the sequencing time and cost. To utilize Sequence Until, \mech integrates a confidence calculation mechanism that evaluates the relative abundance estimations in real-time and fully stops the entire sequencing run if using more reads does not change its estimation. To stop the entire sequencing run for further reads, Run Until can be used to stop the entire sequencing run, which can enable the better utilization of nanopores. We find that Sequence Until can be applied to other mechanisms (e.g., \unc) that can perform real-time relative abundance estimations. Prior work~\cite{weilguny_dynamic_2023} proposes a technique to terminate the sequencing process when species in the sample reach a certain coverage depth. The key difference of Sequence Until is that it reduces the cost of sequencing for relative abundance estimation and is based on our adaptive, accurate, and low-cost confidence calculation during real-time abundance estimation.

We evaluate \mech on three important applications that can benefit from real-time genome analysis: 1)~read mapping, 2)~relative abundance estimation, and 3)~contamination analysis. We compare \mech with the state-of-the-art approaches, \unc and \sig, which can be used with nanopore sequencers that may not be equipped with GPUs, such as the MinION devices. We evaluate \mech, \unc, and \sig in terms of their performance, accuracy, and their estimated benefits in reducing the sequencing time and cost.

This paper provides the following \textbf{key contributions} and  \textbf{major results}:

\begin{itemize}
\item We propose \textbf{\mech}, the \emph{first} mechanism that can efficiently and accurately find the similarities between raw nanopore signals and a reference genome for large genomes without requiring powerful computational resources such as GPUs.
\item We propose the first sampling mechanism that can stop the entire sequencing run for certain applications when an accurate decision can be made without sequencing the entire sample, which we call \textbf{Sequence Until}.
\item We extensively evaluate \mech by comparing it with state-of-the-art approaches, \unc and \sig, on various datasets ranging from small genomes (i.e., genomes with up to 100 million bases) to large genomes (e.g., human genome). Our results show that \mech provides 1)~comparable accuracy to \unc and \sig for small genomes and 2)~significantly better accuracy for large genomes than \unc and \sig.
\item We show that \sig cannot perform real-time genome analysis for large genomes as it cannot match the throughput of nanopores.
\item We provide the open source implementation of \mech and the complete set of scripts to reproduce the results shown in this paper at \release.
\end{itemize}

\section{Methods} \label{sec:methods}
We propose \textbf{\mech}, a mechanism that can efficiently and accurately identify similarities between raw nanopore signals of a read and a large reference genome in real-time (i.e., while the read is sequenced). The raw nanopore signal of each read is a series of electrical current measurements as a strand of DNA passes through a nanopore. The reference genome is a set of strings over the alphabet \texttt{A,C,G,T}. \mechcap provides the mechanisms for generating hash values from both a raw nanopore signal and a reference genome such that similar regions between the two can be efficiently and accurately found by matching their hash values.

\subsection{Overview} \label{subsec:overview}
\begin{figure}[tbh]
  \centering
  \includegraphics[width=0.8\columnwidth]{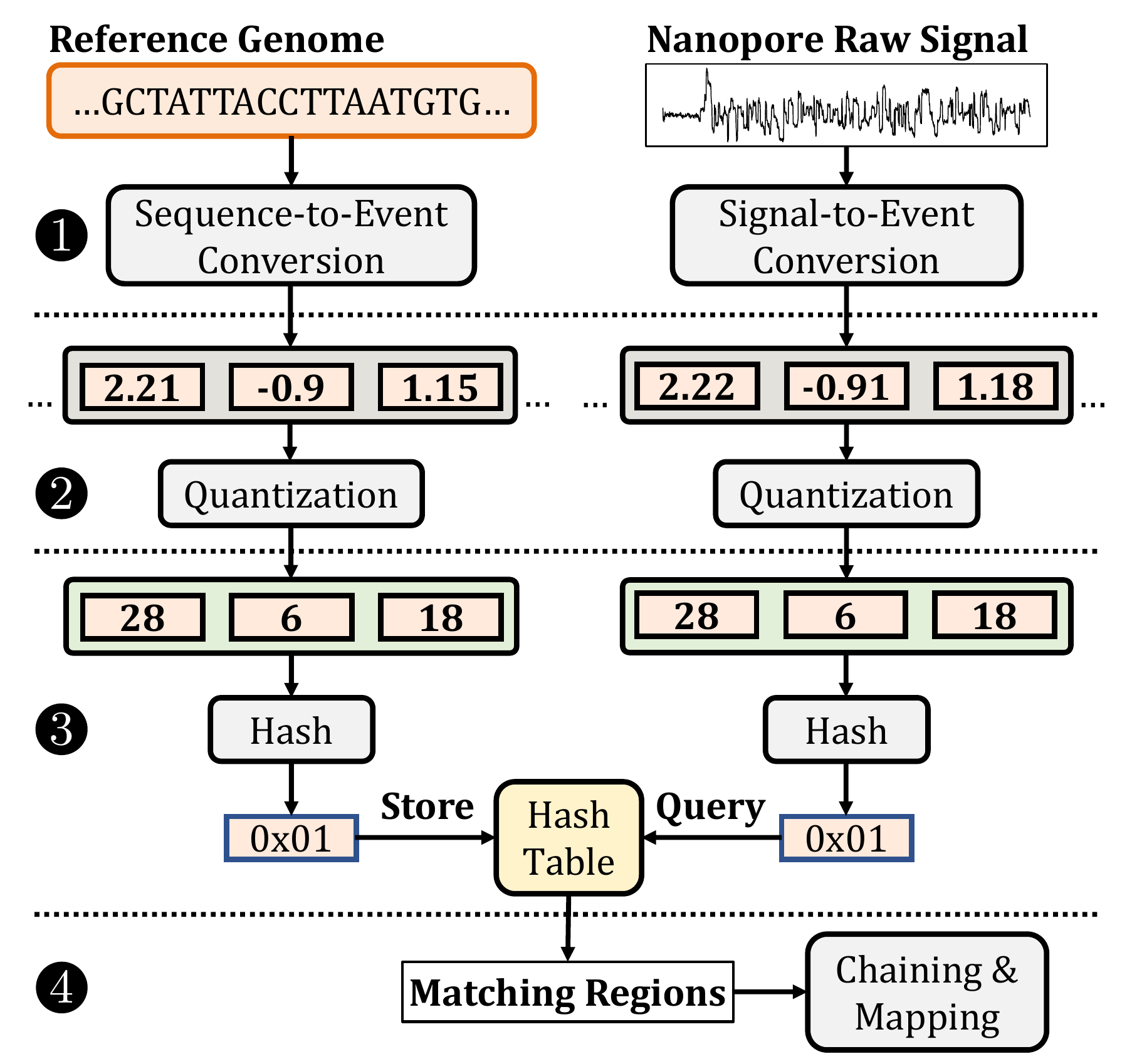}
  \caption{Overview of \mech.}
  \label{fig:overview}
\end{figure}

Figure~\ref{fig:overview} shows the overview of how \mech identifies similarities between raw nanopore signals of a read and a reference genome in four steps. First, \mech pre-processes both 1)~the raw nanopore signal and 2)~the reference genome into values that are comparable to each other. For raw signals, \mech segments the raw signal into non-overlapping regions such that each region is expected to contain a certain amount of signal values that are generated from reading a fixed number $k$ of DNA bases. Each such region is called an \emph{event}~\cite{david_nanocall_2017}. Each event is usually represented with a value derived from the signal values in the segment. For the reference genome, \mech translates each substring of length $k$ (called a \emph{k-mer}) into their \emph{expected} event values based on the nanopore model.

The event values from the reference genome are not directly comparable to the event values from raw nanopore signals due to variability in the current measurements in nanopores generating slightly different event values for the same k-mer~\cite{david_nanocall_2017}. To generate the same values from slightly different events that may contain the same k-mer information, the second step of \mech \emph{quantizes} the event values from a larger set of values into a smaller set. The quantization technique ensures that the event values within a certain range are likely to be assigned to the same quantized value such that the effect of signal variation is alleviated, i.e., the same k-mer is likely assigned the same quantized value.

Due to the nature of nanopores, each event usually represents a very small k-mer of length around $k$=6 bases, depending on the nanopore model~\cite{zhang_real-time_2021}. Such a short k-mer is likely to exist in a large number of locations in the reference genome, making it challenging to efficiently identify the correct one. To make the events more unique (i.e., such that they exist only in a small number of locations in the reference genome), the third step of \mech combines multiple consecutive quantized events into a single hash value. These hash values can then be used to efficiently identify similar regions between raw signals and the reference genome by matching the hash values generated from their events using efficient data structures such as hash tables.

Fourth, to map a raw nanopore signal of a read to a reference genome, \mech uses a chaining algorithm~\cite{zhang_real-time_2021, li_minimap2_2018} that find colinear matching hash values generated from regions that are close to each other both in the reference genome and the raw nanopore signal.

\subsection{Event Generation} \label{subsec:eventgeneration}

Our goal is to translate a reference genome sequence and a raw nanopore signal into comparable values. To this end, \mech converts 1)~each k-mer of the reference genome and 2)~each segmented region of the raw signal into its corresponding event.

\begin{figure}[tbh]
  \centering
  \includegraphics[width=\columnwidth]{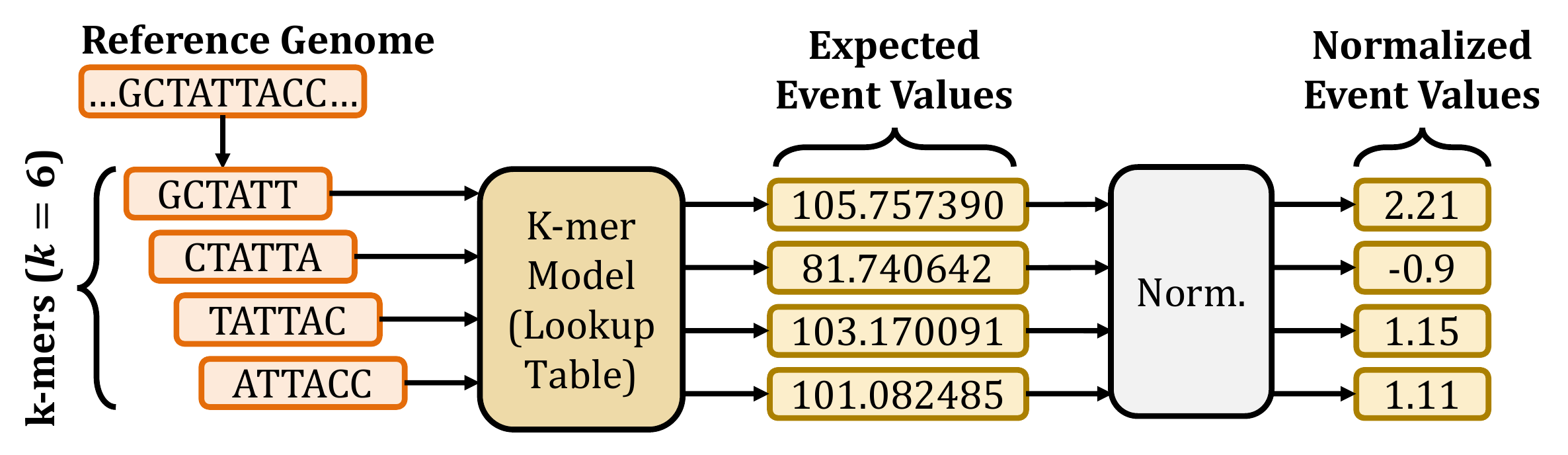}
  \caption{Converting sequences to event values based on the k-mer model of a nanopore.}
  \label{fig:seqtoev}
\end{figure}

\head{Sequence-to-Event Conversion} To convert a reference genome sequence into a form that can be compared with raw nanopore signals, \mech converts the reference genome sequence into event values in three steps, as shown in Figure~\ref{fig:seqtoev}.

First, \mech extracts all k-mers from the reference genome sequence, where $k$ depends on the nanopore. The \emph{k-mer model} of a nanopore\footnote{For many nanopore models, ONT provides the k-mer model including recent R10 and R10.4. These models can also be generated by users~\cite{simpson_detecting_2017}.} includes the information about the \emph{expected} k-mer length of an event and the \emph{expected} average event value for each k-mer based on certain variables affecting the signal outcome of the nanopore's current measurements. 

Second, \mech queries the k-mer model for each k-mer of the reference genome to convert k-mers into their expected event values. Although the k-mer model of a nanopore provides an extensive set of information for each possible k-mer, \mech uses only the mean values of events that provide an average value for the signals in the same event since these mean values provide a sufficient level of meaningful information for comparison with the raw nanopore signals.

Third, \mech normalizes the event values from the same reference genome sequence (e.g., entire chromosome sequence or a contig) by calculating the standard scores (i.e., z-scores) of these events. \mechcap uses these normalized values as event values since the same normalization step is taken for raw signals to avoid certain variables that may affect the range of raw signal amplitudes during sequencing~\cite{zhang_real-time_2021, kovaka_targeted_2021}.

\begin{figure}[tbh]
  \centering
  \includegraphics[width=\columnwidth]{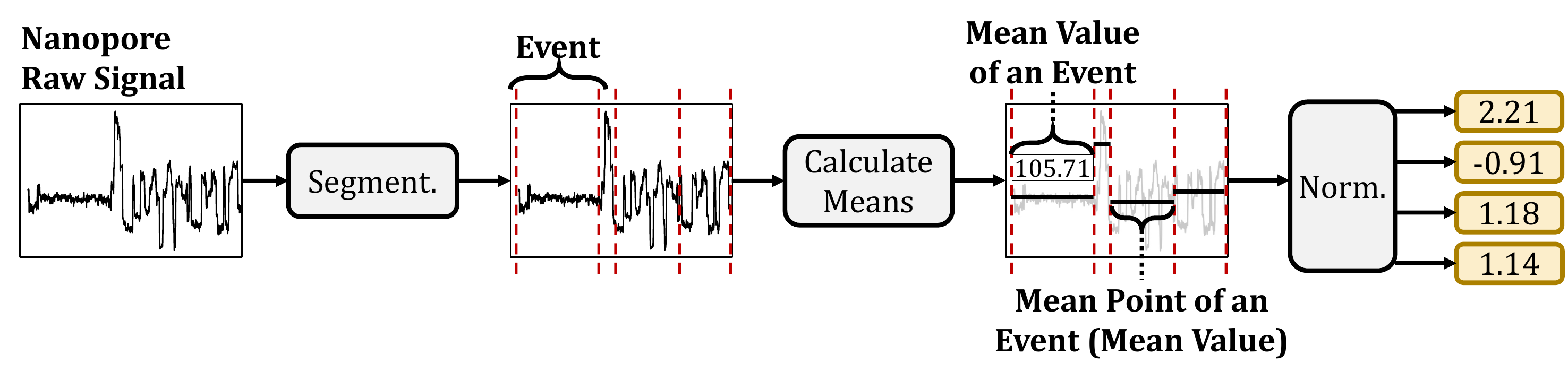}
  \caption{Detecting events from raw signals.}
  \label{fig:sigtoev}
\end{figure}

\head{Signal-To-Event Conversion} Our goal is to accurately convert the series of raw nanopore signals into a set of values where each value corresponds to certain DNA sequences of fixed length $k$, k-mers, and consecutive values differ by one base. To achieve this, \mech converts the raw signals into their corresponding values in three steps, as shown in Figure~\ref{fig:sigtoev}. First, to accurately identify the distinct regions in the raw signal that correspond to a certain k-mer from DNA, \mech performs a segmentation step as described in a basecalling tool, Scrappie, and used by earlier works \unc and \sig. The segmentation step aims to eliminate the factors that affect the speed of the DNA molecules passing through a nanopore, as the speed affects the number of signal measurements taken for a certain amount of bases in DNA. To perform the segmentation step, \mech identifies the boundaries in the signal where the signal value changes significantly compared to the certain amount of previously measured signal values, which indicates a base change in the nanopore. Such boundaries are computed using a statistical test, known as \emph{Welch's t-test}~\cite{ruxton_unequal_2006}, over a rolling window of consecutive signals. \mechcap performs this t-test for multiple windows of different lengths to avoid the variables that cause a change in the number of current measurements due to the varying speed of DNA through a nanopore, known as \emph{skip} and \emph{stay} errors~\cite{david_nanocall_2017}. Signals that fall within the same segment (i.e., between the same measured boundaries) are usually called \emph{events} since each event contains the signals from a reading of a fixed amount of DNA bases, k-mers.

Second, since the number of signals that each event includes is not constant across different events due to the stay and skip errors, \mech generates a single value for each event to quickly avoid these potential errors and other factors that cause variations from reading the same amount of DNA bases. To this end, \mech measures the mean value of the signals that fall within the same segment and uses this mean value for an event.

Third, since the amplitudes of the signal measurements may significantly vary when reading k-mers at different times, \mech normalizes the mean event values using the event values generated from the nanopore within the same certain time interval in a streaming fashion. Although this time interval parameter can be modified in our tool, the default configuration of \mech processes the events of signals generated by the nanopore within one second. For normalization, \mech uses the same z-score calculation that it uses for normalizing the event values generated from reference sequences as described earlier. \mechcap uses these normalized values as event values when comparing with the event values from reference sequences.

\subsection{Quantization of Events} \label{subsec:quantization}

\begin{figure}[tbh]
  \centering
  \includegraphics[width=0.9\linewidth]{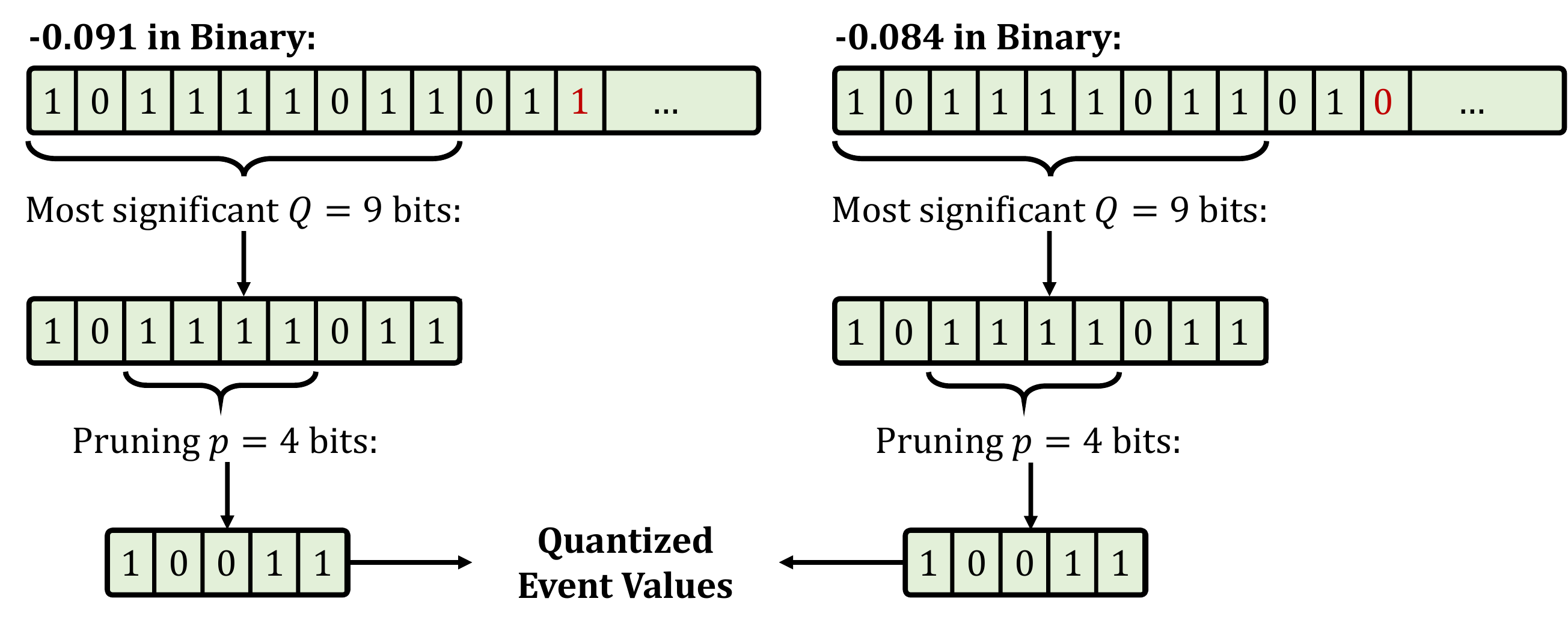}
  \caption{Quantization of two event values.}
  \label{fig:quantization}
\end{figure}

Our goal is to avoid the effects of generating different event values when reading the same k-mer content from nanopores so that we can identify k-mer matches by directly matching events. Although the segmentation and normalization steps explained in Section~\ref{subsec:eventgeneration} can avoid the potential sequencing errors, such as stay and skip errors and significant changes in the current readings at different times, these approaches still do not guarantee to generate \emph{exactly} the same event values when reading the \emph{same} k-mer content. This is because \emph{slight} changes in the normalized event values may occur when reading the same DNA content due to the high sensitivity and stochasticity of nanopores~\cite{david_nanocall_2017}. Thus, it is challenging to generate the same event value for the same k-mer content after the segmentation and normalization steps. Since these event values generated from reading the same k-mer content are expected to be close to each other~\cite{zhang_real-time_2021}, we propose a quantization mechanism that encodes event values so that events with close mean values can have the same quantized value in two steps as shown in Figure~\ref{fig:quantization}.

First, to increase the probability of assigning the same value for similar event values, \mech trims the least significant fractional part of mean values by using only the most significant $Q$ bits of these mean event values from their binary format, which we represent as $E[1,Q]$ for simplicity where $E$ is the event value and $E[1,Q]$ gives the most significant $Q$ bits of $E$. We assume that the mean event values are represented by the standard single-precision floating-point format with the sign, exponent, and fraction bits. This enables \mech to reduce the wide range of floating-point numbers into a smaller range \emph{without} significantly losing from the accuracy such that event values closer to each other can be represented by the same value in the smaller range of values. We can perform this trimming technique without significant sensitivity loss because we observe that these normalized event values mostly use at most six digits from the fractional part of their values, leaving a large number of fractional bits useless.

Second, to avoid using redundant bits that may carry little or no information in the most significant $Q$ bits of an event value, \mech prunes $p$ bits after the most significant two bits of $E[1,Q]$ such that $2+p < Q$ and the resulting \emph{quantized value} is $E[1,2]E[3+p,Q]$. For simplicity, we show the quantized value of $E$ as $E_{Q,p}$. By ignoring these $p$ bits, we effectively pack $Q$ bits into $Q-p$ bits without losing significant information from event values. We can perform such a pruning operation because we observe that the normalized event values are usually in the range $[-3,3]$ such that these $p$ bits provide little information in distinguishing different event values due to the small range of values. We note that these $Q$ and $p$ values are parameters to \mech and can empirically be adjusted based on the required sensitivity and quantization efficiency. This quantization technique enables \mech to assign the same quantized values for a pair of \emph{close} event values, $E$ and $F$, that may be generated from reading the same k-mer such that $E_{Q,p} = F_{Q,p}$ where $|E-F| < \epsilon$ and $\epsilon$ is small enough for two events to represent the same k-mer content. \mechcap always uses the most significant two bits as these two bits consistently carry the most significant information of the normalized event values, including the sign bit.

\subsection{Generating the Hash Values} \label{subsec:hashvalues}

\begin{figure}[tbh]
  \centering
  \includegraphics[width=0.7\columnwidth]{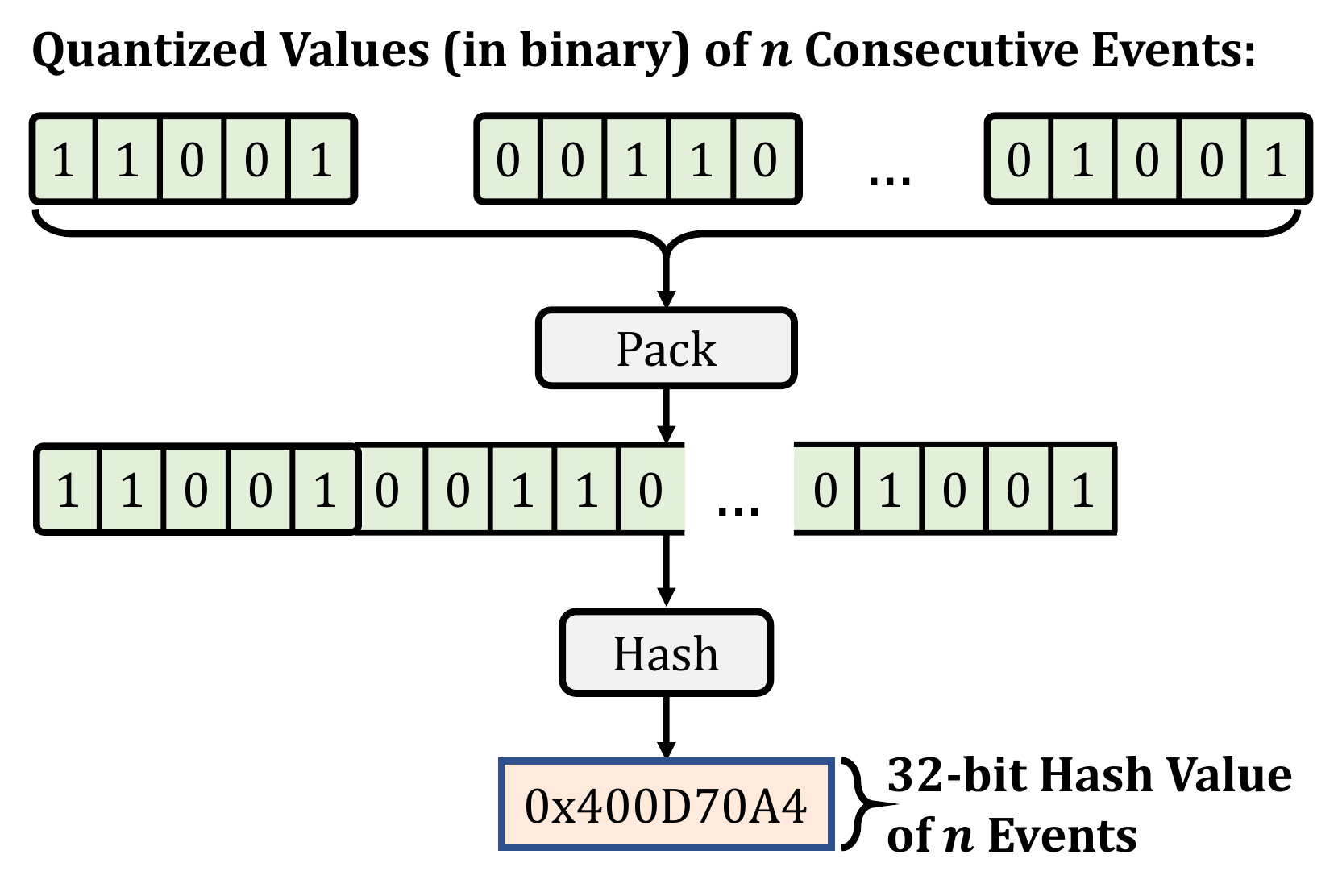}
  \caption{Generating a hash value from $n$ consecutive quantized event values.}
  \label{fig:hashing}
\end{figure}

Our goal is to generate values for \emph{large} regions of raw nanopore signals and reference sequences such that these values can be used to efficiently and accurately identify similarities between raw signals and a reference genome. To this end, \mech generates hash values using quantized values of events in two steps, as shown in Figure~\ref{fig:hashing}. First, to avoid finding a large number of matches, \mech uses the quantized values of $n$ consecutive events to pack them in $n \times (Q-p)$ bits while preserving the order information of these consecutive events. \mechcap uses several consecutive events in a single hash value because matching a single event is likely to generate a larger number of matches for larger genomes as a single event usually corresponds to a k-mer of 6 to 9 bases depending on the nanopore model~\cite{david_nanocall_2017}. It is essential to use several consecutive events to reduce the number of matching regions between raw signals and the reference genome by increasing the region that these consecutive events span.

Second, to efficiently and accurately find matches between large regions of raw signals and a reference genome using a constrained space, \mech uses a low collision hash function to generate a 32-bit hash value from $n \times (Q-p)$ bits of $n$ consecutive quantized event values. Since $n \times (Q-p)$ can be larger than 32, using such a hash function is likely to increase the collision rate for dissimilar regions. To avoid inaccurate similarity identifications due to these incorrect collisions, \mech requires several matches of hash values within close proximity for similarity identification, which we explain next.

\subsection{Seeding and Mapping} \label{subsec:mapping}

To efficiently identify similarities, \mech uses hash values generated from raw nanopore signals and the reference genome in two steps. First, \mech efficiently identifies matching regions between raw nanopore signals and a reference genome by matching their hash values. These hash values used for matching are usually known as \emph{seeds}. Matching seeds enable efficiently finding similar regions between raw nanopore signals and a reference genome. Second, \mech uses the chaining algorithm proposed in Sigmap~\cite{zhang_real-time_2021} to identify the \emph{best} colinear matching seeds that are close to each other in both raw nanopore signal and a reference genome. The region that the \emph{best} chain of seed matches cover is the \emph{mapping position} that \mech identifies as a similar region.

\Copy{R2/1A}{\rev{The chaining algorithm is useful for two reasons. First, the chaining algorithm can tolerate mismatches and indels as it allows including gaps between seed matches, which enables finding similar regions with many seed matches without requiring the entire region to match exactly, as shown in Supplementary Table~\ref{tab:profilinggap}. Second, incorrect seed matches due to collisions or our quantization mechanism that may generate the same quantized value for distinctly dissimilar events are likely to be filtered in the chaining step due to the difficulty of finding colinear seed matches in \rev{highly} dissimilar regions.}} We note that we modify the original chaining algorithm in Sigmap by disabling the distance coefficient as \mech does not calculate the distance between seed matches.

To efficiently map raw signals to a reference genome, \mech provides efficient data structures. To this end, \mech uses hash tables to store the hash values generated from reference genomes (i.e., the indexing step) and efficiently query the same hash table with the hash values generated from the raw signal as the read is sequenced from a nanopore to find positions in the reference genome with matching hash values. \mechcap uses the events in chunks (i.e., collection of events generated within a certain time interval) to find seed matches and perform chaining in a streaming fashion such that the chaining computation from previous chunks (i.e., seed matches) is transferred to the next chunk if the mapping is unsuccessful for the current chunk.

\section{Results} \label{sec:results}
\subsection{Evaluation Methodology} \label{subsec:evaluation}

We implement \mech as a tool for mapping raw nanopore signals to a reference genome. Similar to regular read mapping tools, \mech has two steps to complete the mapping process: 1)~indexing the reference genome and 2)~mapping raw signals. Although indexing is usually a one-time task that can be performed prior to the mapping step, the indexing of \mech can be performed relatively quickly within a few minutes for large genomes \rev{(Supplementary Table~\ref{tab:indexing_resources})}. \mechcap provides the mapping information using a standard pairwise mapping format (PAF). In our implementation, we provide an extensive set of parameters that allow configuring several options to fit \mech for many other applications and nanopore models that we do not evaluate, such as configuring details about the nanopore model (e.g., number of bases per second), number of events that can be included in a single hash value, range of bits to quantize, enabling seeding techniques such as minimizers and fuzzy seed matching. We also provide a default set of parameters that we empirically choose for each common application of real-time genome analysis. These default parameters are set to accurately and efficiently analyze 1)~very small (e.g., viral) genomes, 2)~small and mid-sized genomes (i.e., genomes with less than a few hundred million bases), 3)~large genomes (e.g., genomes with a few billion bases such as a human genome). \Copy{R1/2A}{\rev{We show the details regarding these parameter selections and the versions of tools in Supplementary Tables~\ref{tab:parameters}, \ref{tab:presets}, and \ref{tab:versions}.}}

We evaluate \mech in terms of its performance, peak memory usage, accuracy, and estimated benefits in sequencing time and cost compared to two state-of-the-art tools \unc and \sig. For performance, we evaluate the throughput and overall runtime of each tool in terms of the number of bases they can process per second. Throughput determines if the tool is at least as fast as the speed of DNA passing through a nanopore. For many nanopore models (e.g., R9.4), a DNA strand passes through a pore at around 450 bases per second~\cite{kovaka_targeted_2021, zhang_real-time_2021}. It is essential to provide a throughput higher than the throughput of the nanopore to enable real-time genome analysis. To calculate the throughput, we use the tool that UNCALLED provides, \texttt{UNCALLED pafstats}, which measures the throughput of the tool from the number of bases that the tool processes and the time it takes to process those bases. Although theoretically, it is not possible to exceed the throughput of a nanopore due to the speed of raw signal generation, for comparison purposes, such a limitation is ignored by \texttt{UNCALLED pafstats}. \Copy{R3/2A}{\rev{For overall runtime, we calculate CPU time and real-time using 32 threads. CPU time shows the overall amount of CPU seconds spent running a tool, while real-time shows the overall elapsed (i.e., wall clock) time. All of these tools support multi-threading, where multiple reads can be mapped simultaneously using a single thread for each read. For all of these tools, assigning a larger number of threads enables processing a larger number of reads in parallel, similar to the behavior of nanopore sequencers with hundreds to thousands of pores (i.e., channels). We note that the throughput and mapping time per read values are not affected by the thread counts as 1)~these are measured per read and 2)~single thread performs the mapping of a single read.}}

For accuracy, we evaluate the correctness of the mapping positions that each tool provides when compared to the ground truth mapping positions. To generate the ground truth mapping, we use a read mapping tool, minimap2~\cite{li_minimap2_2018}, to map the basecalled sequences of raw nanopore signals to their corresponding \emph{whole-genome} references. We use \texttt{UNCALLED pafstats} to compare the mapping output of a tool with the ground truth mapping to find the number of true positives or $TP$ (i.e., correct mappings), false positives or $FP$ (i.e., incorrect mappings), and false negatives or $FN$ (i.e., unmapped reads that are mapped in ground truth). Correct and incorrect mappings are identified based on the distance of the mapping positions between ground truth and the tool. To evaluate the accuracy, we calculate the precision ($P = TP/(TP+FP)$), recall ($R = TP/(TP + FN)$) and the $F_1$ ($F_{1} = 2 \times (P \times R)/(P+R)$) values.

\Copy{R2/2B}{\rev{For estimating the benefits in sequencing time and cost of each tool, we calculate the average length of sequenced bases per read when using \unc and \mech and the average number sequenced chunk of signals for \sig and \mech. We compare \mech with \sig in terms of the number of chunks because \sig does not provide the number of bases when a read is unmapped, while both tools provide the number of chunks used when a read is mapped or unmapped. These chunks include a portion of the signal produced by a nanopore within a certain time interval, which is by default set as one second of data for both \mech and \sig. The average length of bases and the number of chunks determine the estimations of how quickly each tool can make a mapping decision to activate Read Until before sequencing the remaining portion of a read, which indicates the potential savings from overall sequencing time and cost.}}

We evaluate \mech, \unc, and \sig for three applications 1)~read mapping, 2)~relative abundance estimation, and 3)~contamination analysis. Read mapping aims to map the raw signals to their corresponding reference genomes. Relative abundance estimation measures the abundance of each genome relative to other genomes in the same sample by mapping raw signals to a given set of reference genomes. Contamination analysis aims to identify if a sample is contaminated with a certain genome (e.g., a viral genome) by mapping raw signals to the reference genome that the sample may be contaminated with. For each tool, we use their default parameter settings in our evaluation.

To evaluate each of these applications, we use real datasets that we list in Table~\ref{tab:dataset}. These datasets include both raw nanopore signals in the FAST5 format and their corresponding basecalled sequences in the FASTA format. \rev{We note that \mech can also use POD5 files.} For relative abundance estimation, we create a mock community using all the read sets from datasets D1 to D5, and the reference genome is the combination of reference genomes used in these datasets. We slightly modify the reference genome we use in the relative abundance estimation such that the sequence IDs in the reference genome provide additional information about the species (e.g., taxonomy IDs) to enable calculating relative abundance in real-time. For contamination analysis, we combine the SARS-CoV-2 read sets (D1) with human read sets (D5) to identify if the combined sample is contaminated with the SARS-CoV-2 sample by mapping raw signals in the combined set to the SARS-CoV-2 reference genome. For all evaluations, we use the AMD EPYC 7742 processor at 2.26GHz to run the tools.

\begin{table}[tbh]
\centering
\caption{Details of datasets used in our evaluation.}
\resizebox{\linewidth}{!}{\begin{tabular}{@{}clrrllr@{}}\toprule
& \textbf{Organism} & \textbf{Reads} & \textbf{Bases}              & \textbf{SRA}       & \textbf{Reference}      & \textbf{Genome}\\
& \textbf{}         & \textbf{(\#)}  & \textbf{(\#)}               & \textbf{Accession} & \textbf{Genome}         & \textbf{Size}  \\\midrule
\multicolumn{7}{c}{Read Mapping} \\\midrule
D1 & \emph{SARS-CoV-2}   & 1,382,016      & 594M                   & CADDE Centre & GCF\_009858895.2   & 29,903 \\\midrule
D2 & \emph{E. coli}      & 353,317        & 2,365M                 & ERR9127551         & GCA\_000007445.1   & 5M \\\midrule
D3 & \emph{Yeast}        & 49,989         & 380M                   & SRR8648503         & GCA\_000146045.2   & 12M\\\midrule
D4 & \emph{Green Algae}  & 29,933         & 609M                   & ERR3237140         & GCF\_000002595.2   & 111M\\\midrule
D5 & \emph{Human HG001}  & 269,507        & 1,584M                 & FAB42260 Nanopore WGS & T2T-CHM13 (v2)     & 3,117M\\\bottomrule
\multicolumn{7}{c}{Relative Abundance Estimation} \\\midrule
\multicolumn{2}{c}{D1-D5} & 2,084,762    & 5,531M                  & D1-D5              & D1-D5              & 3,246M\\\bottomrule
\multicolumn{7}{c}{Contamination Analysis} \\\midrule
\multicolumn{2}{c}{D1, D5} & 1,651,523    & 2,178M                 & D1, D5             & D1                 & 29,903\\\bottomrule
% \multicolumn{7}{l}{For relative abundance estimation and contamination analysis, we combine the reads or}\\
% \multicolumn{7}{l}{reference genomes from the datasets we show in the read mapping section. We use the}\\
\multicolumn{7}{l}{Dataset numbers (e.g., D1-D5) show the combined datasets. Base counts in millions (M).}\\
\end{tabular}}
\label{tab:dataset}
\end{table}

\head{Evaluating Sequence Until} Our goal is to avoid redundant sequencing to reduce sequencing time and cost for relative abundance estimation. We find that the Run Until mechanism can be utilized to \emph{fully} stop the sequencing run when the real-time relative abundance estimation reaches a certain confidence level to achieve accurate estimations, which we call \emph{Sequence Until}. While a similar mechanism is evaluated to enrich the coverage depth of low-abundance species~\cite{weilguny_dynamic_2023} using \ru, we evaluate the potential benefits of Run Until for low-cost relative abundance estimations. We integrate a real-time confidence calculation mechanism in \mech to activate the Sequence Until mechanism in three steps. First, \mech measures the relative abundance estimation after every $n$ reads that can be mapped to a reference genome in real-time. Second, to identify if the recently mapped reads provide substantial changes in the abundance estimations, \mech performs a cross-correlation calculation between the last $w$ estimations. Cross-correlation can identify \emph{outliers} from a set of estimations to identify if the outlier is substantially different than other estimations, which indicates that recent reads can still change the relative abundance estimation, and more reads should be sequenced from the sample. Third, \mech activates Sequence Until by fully stopping the sequencing using Run Until when there are no outliers in the last $w$ estimations, which indicates a convergence to a certain relative abundance estimation, and further sequencing is unlikely to change this estimation. \mechcap provides a set of parameters to adjust these parameters related to Sequence Until.

We evaluate the benefits of Sequence Until by comparing 1)~\mech without Sequence Until and 2)~\mech with Sequence Until in terms of 1)~the difference in the relative abundance estimations and 2)~the estimated benefits in sequencing time and cost. To evaluate Sequence Until in a realistic sequencing environment where reads from different species can be sequenced in a random order, we randomly shuffle the reads in the relative abundance dataset and generate a set of 50,000 reads with a random order of species so that we can simulate this random behavior. We also find that Sequence Until can be applied to other mechanisms. To evaluate the potential benefits of Sequence Until, we simulate the benefits when using UNCALLED with Sequence Until and compare it with \mech.

\subsection{Performance and Peak Memory} \label{subsec:perfmemory}
\begin{figure}[tbh]
  \centering
  \includegraphics[width=\columnwidth]{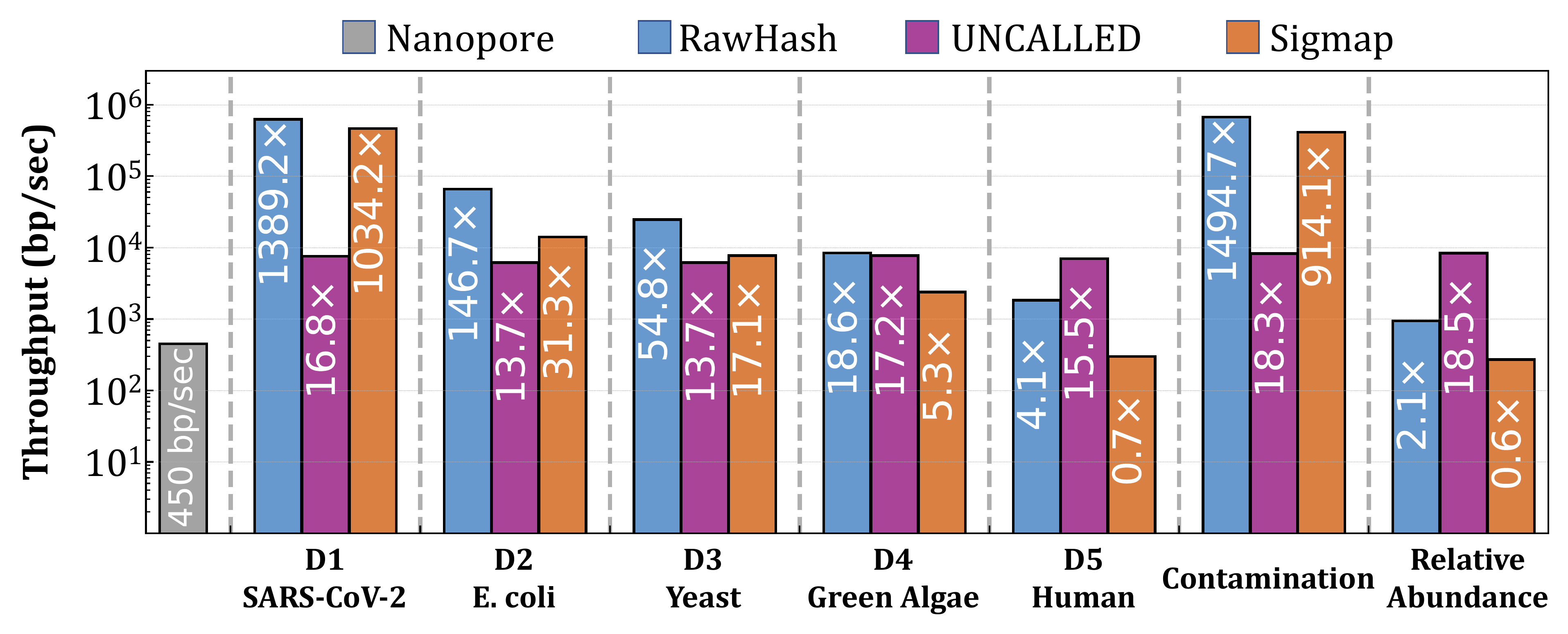}
  \caption{Throughput of each tool. Values inside the bars show the throughput ratio between each tool and a nanopore.}
  \label{fig:throughput}
\end{figure}

Figure~\ref{fig:throughput} shows the throughput of regular nanopores that we use as a baseline and the throughput of the tools when mapping raw nanopore signals to each dataset for read mapping, contamination analysis, and relative abundance estimation. Supplementary Figure~\ref{fig:timeperread} and Supplementary Tables~\ref{tab:indexing_resources} and \ref{tab:mapping_resources} show the mapping time per read, and the computational resources required for indexing and mapping, respectively. We make \rev{six} key observations.
First, \mech and \unc are the \emph{only} tools that can perform real-time genome analysis for large genomes, as they can provide higher throughputs than nanopores for all datasets. \sig \emph{cannot} perform real-time genome analysis for large genomes as it can provide $0.7\times$ and $0.6\times$ throughput of a nanopore for human genome mapping and relative abundance estimations, respectively. \mechcap can achieve high throughput as its seeding mechanism is based on efficiently matching hash values compared to the costly distance calculations that \sig performs for matching seeds, which shows poor scalability for larger genomes.
Second, the throughput of \unc is not affected by the genome size as it provides a near-constant throughput of around $16\times$ for all applications. This is because \unc uses FM-index~\cite{ferragina_opportunistic_2000} and a branching algorithm that provides robust scaling with respect to the reference genome size~\cite{kovaka_targeted_2021}.
\Copy{R1/1D}{\rev{Third, the throughput of \mech decreases with larger genomes as the seeding and chaining steps start taking up a larger fraction of the entire runtime of \mech as shown in Supplementary Table~\ref{tab:profiling}.}}
\rev{Fourth, \mech provides an average throughput \avgthrU and \avgthrS better than \unc and \sig, while providing an average mapping speedup of \avgmtU and \avgmtS per read, respectively. Higher throughput with faster mapping times suggests that the mapping time improvements of \mech are mainly due to its computational efficiency rather than the ability to sequence shorter prefixes of reads than \unc and \sig.}
\rev{Fifth, for indexing, \sig usually requires a larger amount of computational resources in terms of both runtime and peak memory usage.}
\Copy{R3/1A}{\rev{Sixth, for mapping, \unc is the most efficient tool in terms of the peak memory usage as it requires at most 10GB of peak memory while 1)~\mech requires less than 12GB of memory for almost all the datasets and 2)~\sig requires significantly larger memory space than both tools. \mechcap has a larger memory footprint, $\sim52$GB, than \unc for large genomes. Although such large memory requirements for larger genomes can lead to challenges in using \mech for mobile devices with limited computational resources, such a requirement can be mitigated by using more efficient seeding techniques such as minimizers, which we leave as future work.}}
We conclude that \mech provides significant benefits in improving the throughput and performance for the real-time analysis of large genomes while matching the throughput of nanopores.

\subsection{Accuracy} \label{subsec:accuracy}

Table~\ref{tab:accuracy} shows the accuracy results of tools for each dataset and application. We make four key observations. First, \mech provides the best accuracy in terms of precision, recall, and $F_{1}$ values compared to \unc and \sig when mapping reads to large genomes (i.e., the human genome and the relative abundance estimation). \mechcap can efficiently match several events using hash values, which is specifically beneficial in reducing the number of matching regions in large genomes and increasing the specificity due to finding longer matches compared to \unc and \sig.

Second, \mech and \unc can accurately perform contamination analysis while \sig suffers from significantly lower precision and recall values. Due to the nature of a contamination analysis, it is essential to correctly eliminate the genomes other than the contaminating genome (precision) without missing the correct mappings of reads from the contaminating genome (recall). Unfortunately, \sig cannot provide high values in any of these categories, making it significantly unsafe for contamination detection.

Third, the precision of \mech does not drop with the increased length in the reference genome due to the benefits of finding long matches, which provides a higher confidence in read mapping.

Fourth, although \mech does not provide the best accuracy when mapping reads to genomes smaller than the human genome, its accuracy is on par with \unc and \sig for these genomes. \unc and \sig can achieve high recall values as their mechanisms are best optimized for accurately handling matches in relatively smaller genomes with fewer repeats and ambiguous mappings~\cite{zhang_real-time_2021, kovaka_targeted_2021}. We conclude that \mech is the \emph{only} tool that can accurately scale to performing real-time genome analysis for large genomes, especially with significantly high precision rates.

\begin{table}[tbh]
\centering
\caption{Mapping accuracy.}
\resizebox{0.9\linewidth}{!}{\begin{tabular}{@{}llrrr@{}}\toprule
\textbf{Dataset}   &           & \textbf{\unc}   & \textbf{\sig}   & \textbf{\mechcap}  \\\midrule
\multicolumn{5}{c}{Read Mapping} \\\midrule
D1	               & Precision & 0.9547          & \textbf{0.9929} & 0.9868          \\
\emph{SARS-CoV-2}  & Recall    & \textbf{0.9910} & 0.5540          & 0.8735 		 \\
                   & $F_1$     & \textbf{0.9725} & 0.7112          & 0.9267          \\\midrule
D2	               & Precision & 0.9816          & \textbf{0.9842} & 0.9573          \\
\emph{E. coli}     & Recall    & \textbf{0.9647} & 0.9504          & 0.9009 		 \\
                   & $F_1$     & \textbf{0.9731} & 0.9670          & 0.9282          \\\midrule
D3	               & Precision & 0.9459          & 0.9856          & \textbf{0.9862} \\
\emph{Yeast}       & Recall    & \textbf{0.9366} & 0.9123          & 0.8412 		 \\
                   & $F_1$     & 0.9412          & \textbf{0.9475} & 0.9079          \\\midrule
D4	               & Precision & 0.8836          & \textbf{0.9741} & 0.9691          \\
\emph{Green Algae} & Recall    & 0.7778          & \textbf{0.8987} & 0.7015 		 \\
                   & $F_1$     & 0.8273          & \textbf{0.9349} & 0.8139          \\\midrule
D5	               & Precision & 0.4867          & 0.4287          & \textbf{0.8959} \\
\emph{Human HG001} & Recall    & 0.2379          & 0.2641          & \textbf{0.4054} \\
                   & $F_1$     & 0.3196          & 0.3268          & \textbf{0.5582} \\\midrule
\multicolumn{5}{c}{Relative Abundance Estimation} \\\midrule
	              & Precision & 0.7683          & 0.7928           & \textbf{0.9484} \\
D1-D5             & Recall    & 0.1273          & 0.2739           & \textbf{0.3076} \\
                  & $F_1$     & 0.2184          & 0.4072           & \textbf{0.4645} \\\midrule
\multicolumn{5}{c}{Contamination Analysis} \\\midrule
                  & Precision & \textbf{0.9378} & 0.7856           & 0.8733          \\
D1, D5            & Recall    & \textbf{0.9910} & 0.5540           & 0.8735 		   \\
                  & $F_1$     & \textbf{0.9637} & 0.6498           & 0.8734          \\\bottomrule
\multicolumn{5}{l}{\footnotesize Best results are highlighted with \textbf{bold} text.} \\
\end{tabular}
}
\label{tab:accuracy}
\end{table}

\head{Relative Abundance Estimations}
Table~\ref{tab:relativeabundance} shows the relative abundance estimations that each tool makes and the Euclidean distance of their estimation to the ground truth estimation. We make two key observations.
First, we find that \mech provides the most accurate relative abundance estimations in terms of the estimation distance to the ground truth compared to \unc and \sig. This observation correlates with the accuracy results we show in Table~\ref{tab:accuracy} where \mech provides the best overall accuracy for relative estimation, which results in generating the most accurate relative abundance estimations.
Second, although \sig cannot perform real-time relative abundance estimation due to its throughput being lower than a nanopore (Figure~\ref{fig:throughput}), \sig provides accurate estimations that are on par with \mech. This observation shows that while \sig provides mappings with more incorrect positions due to lower precision than \mech (Table~\ref{tab:accuracy}), these reads with incorrect mapping positions are mostly mapped to their correct species.
We conclude that \mech is the \emph{only} tool that can \emph{accurately} be applied to analyze relative abundance estimations while matching the throughput of nanopores at a large-scale based on the prior knowledge of the set of reference genomes to map the reads.

\begin{table}[tbh]
\centering
\caption{Relative abundance estimations.}
\resizebox{\columnwidth}{!}{\begin{tabular}{@{}lrrrrrr@{}}\toprule
              & \multicolumn{6}{c}{Estimated Relative Abundance Ratios} \\\cmidrule{2-7}
\textbf{Tool} & \textbf{\emph{SARS-CoV-2}} & \textbf{\emph{E. coli}} & \textbf{\emph{Yeast}} & \textbf{\emph{Green Algae}} & \textbf{\emph{Human}} & \textbf{Distance} \\\midrule
Ground Truth  & 0.0929                     & 0.4365                  & 0.0698                & 0.1179                      & 0.2828                &  N/A              \\\midrule
\unc          & 0.0026                     & 0.5884                  & 0.0615                & 0.1313                      & 0.2161                & 0.1895            \\\midrule
\sig          & 0.0419                     & 0.4191                  & 0.1038                & 0.0962                      & 0.3390                & 0.0877            \\\midrule
\mechcap      & 0.1249                     & 0.4701                  & 0.0957                & 0.0629                      & 0.2464                & \textbf{0.0847}   \\\bottomrule
\multicolumn{7}{l}{\footnotesize Best results are highlighted with \textbf{bold} text.} \\
\end{tabular}}
\label{tab:relativeabundance}
\end{table}

\subsection{Sequencing Time and Cost} \label{subsec:sequencingcost}
\Copy{R2/2C}{\rev{Our goal is to estimate the benefits that each tool provides in reducing the sequencing time and cost. To this end, we measure the average length of sequenced bases and the average number of sequenced chunks per read. We make two key observations.
First, \mech provides significant benefits in reducing the sequencing time and cost for large genomes (e.g., Green Algae and Human) compared to \unc, as \mech can complete the mapping process per read by using smaller prefixes of reads.
Second, \mech uses on average $1.58 \times$ more chunks compared to \sig when mapping reads, which can proportionally lead to worse sequencing time and cost for \mech compared to \sig.
We conclude that although \unc and \sig provide better advantages in reducing sequencing time and cost for smaller genomes, \mech can provide significant reductions in sequencing time and cost for larger genomes compared to \unc.}}

\begin{table}[tbh]
\centering
\caption{\Copy{R2/2A}{\rev{The average sequenced length of bases and the number of chunks.}}}
\resizebox{\linewidth}{!}{\begin{tabular}{@{}lrrrrr@{}}\toprule
\textbf{Tool}     & \textbf{\emph{SARS-CoV-2}} & \textbf{\emph{E. coli}} & \textbf{\emph{Yeast}} & \textbf{\emph{Green Algae}} & \textbf{\emph{Human}} \\\midrule
\multicolumn{6}{c}{Average sequenced base length per read} \\\midrule
\unc              & \textbf{184.51}            & \textbf{580.52}         & \textbf{1,233.20}     & 5,300.15                    & 6,060.23              \\
\mech             & 513.95                     & 1,376.14                & 2,565.09              & \textbf{4,760.59}           & \textbf{4,773.58}      \\\midrule
\multicolumn{6}{c}{Average sequenced number of chunks per read} \\\midrule
\sig              & \textbf{1.01}              & \textbf{2.11}           & \textbf{4.14}         & \textbf{5.76}               & \textbf{10.40}        \\
\mech             & 1.24                       & 3.20                    & 5.83                  & 10.72                       & 10.70                 \\\bottomrule
\multicolumn{6}{l}{\footnotesize Best results are highlighted with \textbf{bold} text.} \\
\end{tabular}
}
\label{tab:profilingsampling}
\end{table}

\subsection{Benefits of Sequence Until}\label{subsec:sequenceuntil_benefits}

\head{Simulated Sequence Until}
Our goal is to estimate the benefits of implementing the Sequence Until mechanism in \unc and compare it with \mech when they both use Sequence Until under the same conditions. To this end, we use \texttt{shuf} in Linux to randomly shuffle the mapping files that both \mech and \unc generate for relative abundance and extract a certain portion of the randomly shuffled file to identify their relative abundance estimations after $0.01\%$, $0.1\%$, $1\%$, $10\%$, and $25\%$ of the overall reads in the sample are randomly sequenced from nanopores. 

Table~\ref{tab:sequenceuntil} shows the distance of relative abundance estimations after a certain portion of the read is randomly sequenced from nanopores. We make two key observations.
First, \emph{both} \mech and \unc can significantly benefit from Sequence Until by stopping sequencing after processing a smaller portion of the entire sample since their estimations using smaller portions are close to those using the entire set of reads (Table~\ref{tab:relativeabundance}) in terms of their distance to the ground truth. This suggests that many other tools can benefit from Sequence Until as their sensitivity to relative abundance estimations may not significantly change while providing opportunities for reducing the sequencing time and cost up to a certain threshold based on the tool.

Second, \mech can provide more accurate relative abundance estimations when using only $0.1\%$ of the reads than the estimation that \unc provides using the entire set of reads (Table~\ref{tab:relativeabundance}). We conclude that Sequence Until provides significant opportunities in reducing sequencing time and cost while more accurate tools such as \mech can benefit further from Sequence Until by using fewer portions of the entire read set than the portions that less accurate tools would need to achieve similar accuracy.

\begin{table}[tbh]
\centering
\caption{Relative abundance with simulated Sequence Until.}
\resizebox{\columnwidth}{!}{\begin{tabular}{@{}lrrrrrr@{}}\toprule
                 & \multicolumn{6}{c}{Estimated Relative Abundance Ratios} \\\cmidrule{2-7}
\textbf{Tool}    & \textbf{\emph{SARS-CoV-2}} & \textbf{\emph{E. coli}} & \textbf{\emph{Yeast}} & \textbf{\emph{Green Algae}} & \textbf{\emph{Human}} & \textbf{Distance} \\\midrule
Ground Truth     & 0.0929                     & 0.4365                  & 0.0698                & 0.1179                      & 0.2828                &  N/A              \\\midrule
\unc ($25\%$)    & 0.0026                     & 0.5890                  & 0.0613                & 0.1332                      & 0.2139                & 0.1910          \\
\mechcap ($25\%$)   & 0.0271                     & 0.4853                  & 0.0920                & 0.0786                      & 0.3170                & \textbf{0.0995} \\\midrule
\unc ($10\%$)    & 0.0026                     & 0.5906                  & 0.0611                & 0.1316                      & 0.2141                & 0.1920          \\
\mechcap ($10\%$)   & 0.0273                     & 0.4869                  & 0.0963                & 0.0772                      & 0.3124                & \textbf{0.1004} \\\midrule
\unc ($1\%$)     & 0.0026                     & 0.5750                  & 0.0616                & 0.1506                      & 0.2103                & 0.1836          \\
\mechcap ($1\%$)    & 0.0259                     & 0.4783                  & 0.0987                & 0.0882                      & 0.3088                & \textbf{0.0928} \\\midrule
\unc ($0.1\%$)   & 0.0040                     & 0.4565                  & 0.0380                & 0.1910                      & 0.3105                & 0.1242          \\
\mechcap ($0.1\%$)  & 0.0212                     & 0.5045                  & 0.1120                & 0.0810                      & 0.2814                & \textbf{0.1136} \\\midrule
\unc ($0.01\%$)  & 0.0000                     & 0.5551                  & 0.0000                & 0.0000                      & 0.4449                & 0.2602          \\
\mechcap ($0.01\%$) & 0.0906                     & 0.6122                  & 0.0000                & 0.0000                      & 0.2972                & \textbf{0.2232} \\\bottomrule
\multicolumn{7}{l}{\footnotesize Percentages show the portion of the overall reads used. Best results are highlighted with \textbf{bold} text.} \\
\end{tabular}}
\label{tab:sequenceuntil}
\end{table}

\head{Sequence Until with \mech}
Our goal is to evaluate Sequence Until when used in real-time with \mech for relative abundance estimation. Table~\ref{tab:sequenceuntil_real} shows the relative abundance estimations that \mech makes with and without Sequence Until. We note that the estimations we show for \mech in Table~\ref{tab:sequenceuntil_real} are different than the estimations in Table~\ref{tab:relativeabundance} since we randomly subsample the reads in the relative abundance estimation dataset, as explained in Section~\ref{subsec:evaluation}. We make two key observations. First, we observe that the distance between the relative abundance estimations between these two configurations of \mech is substantially low. This indicates that our outlier detection mechanism can accurately detect the convergence to the relative abundance estimations without using a full set of reads. Second, Sequence Until enables accurately stopping the entire sequencing after processing $7\%$ of the reads in the entire set without substantially sacrificing accuracy. We conclude that Sequence Until has the potential to significantly reduce the sequencing time and cost by using only fewer reads from a sample while producing accurate results.

\begin{table}[tbh]
\centering
\caption{Relative abundance with Sequence Until.}
\resizebox{\columnwidth}{!}{\begin{tabular}{@{}lrrrrrr@{}}\toprule
                       & \multicolumn{6}{c}{Estimated Relative Abundance Ratios in 50,000 Random Reads} \\\cmidrule{2-7}
\textbf{Tool}          & \textbf{\emph{SARS-CoV-2}} & \textbf{\emph{E. coli}} & \textbf{\emph{Yeast}} & \textbf{\emph{Green Algae}} & \textbf{\emph{Human}} & \textbf{Distance} \\\midrule
\mechcap ($100\%$)     & 0.0270                     & 0.3636                  & 0.3062                & 0.1951                      & 0.1081                & N/A          \\\midrule
\mechcap +             & 0.0283                     & 0.3539                  & 0.3100                & 0.1946                      & 0.1133                & 0.0118          \\
Sequence Until ($7\%$) &                            &                         &                       &                             &                       &                 \\\bottomrule
\multicolumn{7}{l}{\footnotesize Percentages show the portion of the overall reads used.} \\
\end{tabular}}
\label{tab:sequenceuntil_real}
\end{table}
\section{Discussion} \label{sec:discussion}

We discuss the benefits we expect \mech can immediately make, the limitations of \mech, and future work. We envision that \mech can be useful mainly for two directions. First, \mech provides a low-cost solution for analyzing large genomes in real-time. Such an analysis can be significantly useful when using nanopore sequencers with limited computational resources to enable portable real-time genome analysis at a large scale.

Second, we expect that \mech can also be useful for genome analysis that does not require real-time solutions by reducing the time and energy that further steps in genome analysis may require. One of the immediate steps after generating raw nanopore signals is their translation to their corresponding DNA bases as sequences of characters with a computationally-intensive step, \emph{basecalling}. Basecalling approaches are usually computationally costly and consume significant energy as they use complex deep learning models~\cite{singh2022framework, mao2022genpip}. Although we do not evaluate in this work, we expect that \mech can be used as a low-cost filter~\cite{cavlak2022targetcall} to eliminate the reads that are unlikely to be useful in downstream analysis, which can reduce the overall workload of basecallers and downstream analysis.

\head{Future work} We find three key directions for future work. First, we find that our efficient hash-based similarity identification mechanism can be used to efficiently find overlaps between signals as the reads are sequenced in real-time. Although we observe that our indexing technique is efficient in terms of the amount it requires to construct an index even for large genomes, such an overlapping technique requires substantially more optimized indexing methods and techniques that can efficiently find overlaps as more reads are sequenced and \emph{evolves} the index. Finding overlaps between signals can be beneficial in 1)~providing enriched information to basecallers to increase their accuracy and 2)~identifying \emph{redundant} signals that fully overlap with already sequenced reads in an effort for generating assemblies from signals.

Second, since \mech generates hash values for matching similar regions, it provides opportunities to use the hash-based seeding techniques that are optimized for identifying sequence similarities accurately without requiring large memory space, such as minimizers~\cite{roberts_reducing_2004, li_minimap2_2018}, spaced seeds~\cite{ma_patternhunter_2002}, syncmers~\cite{edgar_syncmers_2021}, strobemers~\cite{sahlin_effective_2021}, and fuzzy seed matching as in BLEND~\cite{firtina_blend_2023}. \Copy{R1/1E}{Although we do not evaluate in this work, we implement the minimizer seeding technique in \mech. Our initial observation motivates us that future work can exploit these seeding techniques with slight modifications in their seeding mechanisms to significantly improve the performance of certain applications without reducing the accuracy.}

Third, we find that \mech can also benefit from a GPU implementation as its low-cost and accurate implementation can effectively be scaled to nanopore sequencers that include thousands of nanopores such that these pores can be analyzed in parallel with an efficient GPU implementation, which we leave as future work.
\section{Related Work} \label{sec:relatedwork}
\rev{To our knowledge, \emph{\mech} is the first mechanism to efficiently and accurately perform real-time analysis of raw nanopore signals for large genomes. We discuss related work in 1)~basecalling, 2)~accelerating genome analysis after the basecalling step, and 3)~real-time genome analysis with limited computational resources.

\head{Basecalling} Deep learning-based models are utilized by modern basecallers to considerably enhance the precision of identifying a nucleotide base from raw signals compared to traditional non-deep learning-based basecallers~\cite{urnet_zhang2020nanopore, dias2019artificial, amarasinghe2020opportunities, senol_cali_nanopore_2019, rang2018squiggle, singh2022framework}. Deep learning models can successfully basecall genomes due to the developments and advancements in their architecture, which enables them to model and accurately recognize spatial characteristics in the raw data. Many basecallers have been proposed using modern deep learning-based architectures~\cite{bonito, konishi_halcyon_2021, huang2020sacall, xu2021fast, boza_deepnano_2017, Guppy, perevsini2021nanopore, lv_end--end_2020, zeng_causalcall_2020, yeh_msrcall_2022} However, the use of complex deep learning models makes basecalling slow and memory-hungry, bottlenecking all genomic analyses that depend on it~\cite{singh2022framework}. Recent works focus on developing methods to speed up the basecalling process. One approach to basecalling acceleration is to use specialized hardware, such as field-programmable gate arrays (FPGAs)~\cite{wu2018fpga, ramachandra_ont-x_2021, hammad_scalable_2021, wu_fpga_2022, wu_fpga-accelerated_2020} or processing-in-memory (PIM)~\cite{mao2022genpip, lou2020helix, lou2018brawl}, to perform the basecalling computations. These specialized hardware devices can perform many calculations in parallel, allowing for significant speedups in the basecalling process. Another approach is to use machine learning-based compression techniques to improve the performance of the basecalling process. RUBICON~\cite{singh2022framework} provides a framework to develop hardware-optimized basecallers using neural architecture search~\cite{zoph2016neural}, knowledge distillation~\cite{bucilua2006model}, and pruning~\cite{lecun1989optimal}. Dorado~\cite{dorado}, a basecaller by ONT, uses quantization~\cite{gray1998quantization} to reduce the bit-width precision at which neural network calculations are performed.  All the above works accelerate the basecalling step without eliminating the wasted computation in basecalling. TargetCall~\cite{cavlak2022targetcall} proposes a pre-basecalling filter that eliminates the wasted computation in basecalling by leveraging the observation that the majority of reads are discarded after basecalling. However, \mech is different from these works as its goal is to perform real-time analysis of raw signals without performing the computationally-intensive basecalling step. 

\head{Accelerating the genome analysis after basecalling} There are several works that aim to accelerate the entire genome analysis pipeline by accelerating one or multiple steps in the pipeline after basecalling the raw nanopore signals~\cite{alser_molecules_2022, alser_technology_2021}. These works accelerate the pre-alignment filtering and read classification~\cite{xin2013accelerating, xin2015shifted, alser2017gatekeeper, kim2018grim, kaplan2018rassa, alser2019shouji, alser2020sneakysnake, singh2021fpga, alser2017magnet, bingol2021gatekeeper, khalifa_filtpim_2021, mansouri2022genstore, shahroodi_demeter_2022}, chaining~\cite{guo_hardware_2019, sadasivan_accelerating_2022}, read mapping and sequence alignment~\cite{chen2013hybrid, khatamifard2017non, turakhia2018darwin, s_d_goenka_segalign_2020, nag2019gencache, aguado-puig_accelerating_2022, aguado-puig_wfa-gpu_2022, haghi_fpga_2021, cali2020genasm, lindegger2022algorithmic, lindegger2022scrooge, cali_segram_2022, fujiki2018genax, madhavan2014race, cheng2018bitmapper2, houtgast2018hardware, houtgast2017efficient, zeni2020logan, ahmed2019gasal2, nishimura2017accelerating, de2016cudalign, liu2015gswabe, liu2013cudasw++, liu2009cudasw++, liu2010cudasw++, wilton2015arioc, goyal2017ultra, chen2016spark, laguna2020seed, chen2014accelerating, chen2021high, fujiki2020seedex, banerjee2018asap, fei2018fpgasw, waidyasooriya2015hardware, li2021pim, chen2015novel, rucci2018swifold, diab2022framework, zokaee2019finder, angizi2020exploring, diab2022high, huangfu_radar_2018, chowdhury_dna_2020, li2021pipebsw, wu2019fpga, yan_accel-align_2021, daily_parasail_2016, kalikar_accelerating_2022, marco-sola_fast_2021, kaplan2017resistive, khatamifard2021genvom, chen2020parc, gupta2019rapid, zokaee2018aligner, eizenga_improving_2022, firtina_aphmm_2022, marco-sola_optimal_2022} steps. Although these works can significantly improve the performance of the genome analysis pipeline, unlike \mech, these works cannot perform real-time genome analysis while the raw nanopore signals are generated from nanopore sequencers.

\head{Real-time analysis of raw nanopore signal} Several works perform real-time genome analysis of raw nanopore signals by utilizing adaptive sampling~\cite{kovaka_targeted_2021, zhang_real-time_2021, payne_readfish_2021, edwards_real-time_2019, dunn2021squigglefilter, bao_squigglenet_2021, shih_efficient_2022, sadasivan_rapid_2023}.
SquiggleFilter~\cite{dunn2021squigglefilter} uses an ASIC accelerator that quickly filters non-related raw electrical signals before basecalling for viral detection. HARU~\cite{shih_efficient_2022} is an FPGA accelerator that accelerates real-time selective genome sequencing on resource-constrained devices for detecting viral genomes. \mechcap differs from these works as it does not require specialized hardware design and can scale to analyze large genomes while matching the throughput of nanopores.

SquiggleNet~\cite{bao_squigglenet_2021}, DeepSelectNet~\cite{senanayake_deepselectnet_2023}, and RawMap~\cite{sadasivan_rapid_2023} require training with machine learning techniques using sequencing reads as training data without using reference genomes. These works train their models to classify raw nanopore signals without mapping them to the reference genome, which is different than \mech as it maps raw signals to a reference genome. These works often require retraining and reconfiguring the neural network model and architectures. Although such classification approaches can provide high accuracy in labeling reads as target or non-target reads based on a target genome of interest, it can be challenging to easily perform real-time analysis with high accuracy without retraining or reconfiguring these models. \mechcap is different than these works as it can map reads to any reference genome using easily configurable parameter settings.}

\Copy{R4/1A}{\rev{ReadFish~\cite{payne_readfish_2021} and ReadBouncer~\cite{ulrich_readbouncer_2022} can scale to mapping reads to large genomes such as a human genome using GPUs or CPUs (e.g., DeepNano-Blitz~\cite{boza_deepnano-blitz_2020}) for performing basecalling. Similar to ReadFish and ReadBouncer, RUBRIC~\cite{edwards_real-time_2019} use a basecalling approach followed by mapping the basecalled raw signals to analyze raw nanopore signals in real-time. These basecalling approaches are optimized to use the \emph{entire} raw nanopore signal of a read rather than the portions of raw signals produced in real-time, which can be challenging in generating an accurate mapping with a small number of basecalled signals~\cite{kovaka_targeted_2021, zhang_real-time_2021}. \mech differs from ReadFish and ReadBouncer as it does not require powerful computational resources for basecalling, which may not be immediately available for portable sequencers such as ONT MinION. \mechcap can directly and accurately map a small number of raw signals (e.g., signals produced in one second) to a reference genome without basecalling them.

We note that ReadFish and ReadBouncer use an interface, \emph{MinKNOW}, required for adaptive sampling in nanopore sequencing. MinKNOW enables tools to analyze the raw nanopore signals and perform adaptive sampling by using functionalities such as Read Until. However, the throughput of these tools using MinKNOW \emph{cannot} exceed the throughput of a nanopore sequencer. Thus, it becomes challenging to fairly compare these tools with the other tools, such as \mech and \sig, for two reasons. First, the throughput of \mech and \sig can be significantly larger than the throughput of a nanopore (Figure~\ref{fig:throughput}) due to the lack of support for the MinKNOW interface in these tools. Second, the parameters of \mech are empirically chosen to provide the best throughput and accuracy without the potential effects of MinKNOW. It is likely that the accuracy of \mech can improve while providing the same throughput as a nanopore sequencer. We leave the implementation of MinKNOW for \mech as future work as well as the comparison of \mech with ReadFish and ReadBouncer.}}

\Copy{R3/3A}{\rev{\unc~\cite{kovaka_targeted_2021} and \sig~\cite{zhang_real-time_2021} are the most relevant works to \mech. These works map raw nanopore signals to a reference genome without using powerful computational resources (e.g., GPUs), which can be directly used with portable nanopore sequencers. \unc detects events from raw signals, and the probability of k-mers that each event can represent is calculated using k-mer models. \unc identifies the sequence of matching k-mers between the most probable k-mers of events and a reference genome using an FM-index~\cite{ferragina_opportunistic_2000}. However, it becomes challenging to accurately identify the matching regions with such a probabilistic model from a large number of matches as the genome size increases~\cite{kovaka_targeted_2021} (Table~\ref{tab:accuracy}). Thus, \unc is highly accurate for small genomes (e.g., \emph{E. coli} and \emph{Yeast} genomes) due to the smaller number of probabilistic matches in the reference genome that can be identified accurately.

\sig can map raw nanopore signals to genomes larger than the \emph{Yeast} genome (e.g., \emph{Green Algae} with around 100M bases). To achieve this, \sig converts the k-mers of the reference genome into events and matches the events between raw nanopore signals with the events of the reference genome. Since events are not necessarily identical when reading the same DNA content, it is challenging to find accurate matches between them due to the signal variations we discuss in Section~\ref{subsec:eventgeneration}. To address this challenge, \sig creates a vector from each $n$ consecutive events (i.e., $n$-dimensional vector space) from the reference genome (i.e., the indexing step) and measures the Euclidean distance between these vectors and the vectors generated from raw nanopore signals (i.e., the mapping step) using a k-d tree structure. Although the distance between vector of events generated from similar regions is close, such a distance calculation is computationally \emph{costly} and suffers from the \emph{curse of dimensionality} that fundamentally prevents accurately and efficiently increasing the number of events within a single vector, which makes it ineffective for larger genomes.

\mech is different from \unc and \sig as it identifies similarities between a reference genome and a raw nanopore signal by efficiently and accurately matching the hash values generated from them without using 1)~probabilistic model as proposed in \unc that can be inaccurate for large genomes or costly distance calculations.}}

\section{Conclusion} \label{sec:conclusion}
We propose \mech, a novel mechanism that provides a low-cost and accurate approach for real-time genome analysis for large genomes. \mechcap can efficiently and accurately perform real-time analysis of raw nanopore signals to identify similarities between the signals and a reference genome in real-time at a large-scale (e.g., whole-genome analysis for human or communities with multiple samples). To efficiently and accurately identify similarities, \mech 1)~generates events from both raw signals and the reference genome, 2)~quantizes the events into values such that slightly different events that correspond to the same DNA content can have the same value, and 3)~generates hash values from multiple events to efficiently find matching regions between raw signals and a reference genome using hash values with efficient data structures such as hash tables. We compare \mech with the state-of-the-art approaches, \unc and \sig, on three important applications in terms of their performance, accuracy, and estimated benefits in reducing sequencing time and cost. Our results show that 1)~\mech is the \emph{only} tool that can be accurately applied to analyze raw nanopore signals at large-scale, 2)~provides \avgthrU and \avgthrS better average throughput, and 3)~can map reads \avgmtU and \avgmtS faster than \unc and \sig, respectively.

\section*{Acknowledgments}

We thank the SAFARI Research Group members for their valuable feedback and the stimulating intellectual and scholarly environment they provide. We thank the anonymous reviewers of ISMB/ECCB 2023. We acknowledge the generous gifts of our industrial partners, including Intel and VMware. This work is also partially supported by the European Union’s Horizon programme for research and innovation [101047160 - BioPIM] and the Swiss National Science Foundation (SNSF) [200021\_213084].

%%%%%%% -- PAPER CONTENT ENDS -- %%%%%%%%

%%%%%%%%% -- BIB STYLE AND FILE -- %%%%%%%%
\balance

\setstretch{0.7}
\bibliographystyle{IEEEtran}
{\footnotesize \bibliography{main}}
%%%%%%%%%%%%%%%%%%%%%%%%%%%%%%%%%%%%
\setstretch{1}
\onecolumn
\setcounter{secnumdepth}{3}
\clearpage
\begin{center}
\textbf{\LARGE Supplementary Material for\\ \ltitle}
\end{center}
%%%%%%%%%% Merge with supplemental materials %%%%%%%%%%
%%%%%%%%%% Prefix a "S" to all equations, figures, tables and reset the counter %%%%%%%%%%
\setcounter{section}{0}
\setcounter{equation}{0}
\setcounter{figure}{0}
\setcounter{table}{0}
\setcounter{page}{1}
\makeatletter
\renewcommand{\theequation}{S\arabic{equation}}
\renewcommand{\thetable}{S\arabic{table}}
\renewcommand{\thefigure}{S\arabic{figure}}
\renewcommand{\thesection}{S\arabic{section}}
\renewcommand{\thetheorem}{S\arabic{theorem}}

\renewcommand\rev[1]{{\color{black}{#1}}}

\setstretch{0.95}
\newcommand{\TextUnderscore}{\rule{.4em}{.4pt}}
\section{Profiling \mech} \label{sec:profiling}
\subsection{Profiling the Performance}\label{subsec:profilingper}
\Copy{R1/1A}{\rev{To analyze the potential bottlenecks and computational overheads in \mech, we measure the runtime of five steps in \mech that mainly make up the entire mapping of a read: 1)~I/O operations, 2)~signal-to-event conversion (Section~\ref{subsec:eventgeneration}), 3)~sketching (i.e., quantizing, packing, and hashing events to use them as seeds as described in Sections~\ref{subsec:quantization} and \ref{subsec:hashvalues}), 4)~seeding (Section~\ref{subsec:mapping}), and 5)~chaining (Section~\ref{subsec:mapping}). 

Supplementary Table~\ref{tab:profiling} shows the breakdown analysis of each step in \mech on various datasets and the overall percentage that both seeding and chaining steps take over the entire runtime of \mech. We make two key observations. First, we find that the chaining step is the main computational bottleneck in \mech for all datasets. This is mainly because chaining requires several more computationally costly calculations (e.g., dynamic programming-based computations and sorting) than the other steps. Second, the chaining and seeding steps combined take a larger fraction of the overall runtime as the size of the genome increases (columns in Supplementary Table~\ref{tab:profiling} show increasing genome size from left to right). As the search space increases with larger genomes, the index stores a larger number of seeds for a reference genome. This can increase 1)~the time for finding seed matches due to the way hash table structure is implemented in \mech similar to minimap2~\citesupp{supp_li_minimap2_2018} and 2)~the number of seed matches per read. The increased number of seed matches can then also increase the time spent in the chaining step as the number of \emph{anchors} (i.e., seed matches in chains) and chains to process increase proportionally.}}

\begin{table}[tbh]
\centering
\caption{\Copy{R1/1B}{\rev{Runtime of the steps in \mech on various datasets.}}}
\resizebox{0.5\linewidth}{!}{\begin{tabular}{@{}lrrrrr@{}}\toprule
                 & \multicolumn{5}{c}{Fraction of entire runtime (\%)} \\\cmidrule{2-6}
\textbf{Tool}      & \textbf{\emph{SARS-CoV-2}} & \textbf{\emph{E. coli}} & \textbf{\emph{Yeast}} & \textbf{\emph{Green Algae}} & \textbf{\emph{Human}} \\\midrule
File I/O & 0.00 & 0.00 & 0.00 & 0.00 & 0.00 \\
Signal-to-Event & 21.75 & 1.86 & 1.01 & 0.53 & 0.02 \\
Sketching & 0.74 & 0.06 & 0.04 & 0.03 & 0.00 \\
Seeding & 3.86 & 4.14 & 3.52 & 6.70 & 5.39 \\
Chaining & 73.50 & 93.92 & 95.42 & 92.43 & 94.46 \\\midrule
Seeding + Chaining & 77.36 & 98.06 & 98.94 & 99.14 & 99.86 \\\bottomrule
\end{tabular}}
\label{tab:profiling}
\end{table}

\subsection{Profiling the Chaining Gap Sensitivity}\label{subsec:profilinggap}

\Copy{R2/1B}{\rev{Supplementary Table shows the average length of the gap between a pair of anchors that \mech finds when mapping raw nanopore signals in various datasets. Read and reference anchors show the average gap length between a pair of anchors found in reads and the reference genome, respectively. We find that the chaining algorithm can tolerate a large number of mismatches and indels especially for larger genomes without significantly sacrificing the mapping accuracy (Table~\ref{tab:accuracy}).}}

\begin{table}[tbh]
\centering
\caption{\Copy{R2/1C}{\rev{The average gap length between a pair of anchors in reads and a reference genome.}}}
\resizebox{0.5\linewidth}{!}{\begin{tabular}{@{}lrrrrr@{}}\toprule
\textbf{Tool}     & \textbf{\emph{SARS-CoV-2}} & \textbf{\emph{E. coli}} & \textbf{\emph{Yeast}} & \textbf{\emph{Green Algae}} & \textbf{\emph{Human}} \\\midrule
Read Anchors      & 25.37                      & 42.92                   & 79.31                 & 148.58                      & 200.06\\
Reference Anchors & 20.06                      & 33.39                   & 67.55                 & 127.03                      & 165.44\\\bottomrule
\end{tabular}
}
\label{tab:profilinggap}
\end{table}

\section{Runtime and Peak Memory Usage} \label{sec:cpuandmemory}

\Copy{R3/1B}{\rev{Supplementary Tables~\ref{tab:indexing_resources} and \ref{tab:mapping_resources} show the computational resources required by each tool during the indexing and mapping steps, respectively. To measure the required computational resources, we collect CPU time, real time, and peak memory usage of each tool for all the datasets. To collect these results, we use \texttt{time -v} command in Linux.

CPU time shows the total user and system time. The real time shows the overall elapsed (i.e., wall clock) time while the application is running. Peak memory usage shows the maximum resident set size in the main memory that the application required to complete its task. We use 32 threads for all applications.}}

\begin{table*}[tbh]
\centering
\caption{\Copy{R3/1C}{\rev{Computational resources required in the indexing step of each tool.}}}
\resizebox{0.8\linewidth}{!}{\begin{tabular}{@{}lrrrrrrr@{}}\toprule
\textbf{Tool} & \textbf{\emph{Contamination}} & \textbf{\emph{SARS-CoV-2}} & \textbf{\emph{E. coli}} & \textbf{\emph{Yeast}} & \textbf{\emph{Green Algae}} & \textbf{\emph{Human}} & \textbf{\emph{Relative Abundance}} \\\midrule
\multicolumn{8}{c}{CPU Time (sec)} \\\midrule
\unc & 8.72 & 9.00 & 11.08 & 18.62 & 285.88 & 4,148.10 & 4,382.38 \\
\sig & 0.02 & 0.04 & 8.66 & 24.57 & 449.29 & 36,765.24 & 40,926.76 \\
\mech & 0.18 & 0.13 & 2.62 & 4.48 & 34.18 & 1,184.42 & 788.88 \\\midrule
\multicolumn{8}{c}{Real time (sec)} \\\midrule
\unc & 1.01 & 1.04 & 2.67 & 7.79 & 280.27 & 4,190.00 & 4,471.82 \\
\sig & 0.13 & 0.25 & 9.31 & 25.86 & 458.46 & 37,136.61 & 41,340.16 \\
\mech & 0.14 & 0.10 & 1.70 & 2.06 & 15.82 & 278.69 & 154.68 \\\midrule
\multicolumn{8}{c}{Peak memory (GB)} \\\midrule
\unc & 0.07 & 0.07 & 0.13 & 0.31 & 11.96 & 48.44 & 47.81 \\
\sig & 0.01 & 0.01 & 0.40 & 1.04 & 8.63 & 227.77 & 238.32 \\
\mech & 0.01 & 0.01 & 0.35 & 0.76 & 5.33 & 83.09 & 152.80 \\\bottomrule
\end{tabular}
}
\label{tab:indexing_resources}
\end{table*}

\begin{table*}[tbh]
\centering
\caption{\Copy{R3/1D}{\rev{Computational resources required in the mapping step of each tool.}}}
\resizebox{0.8\linewidth}{!}{\begin{tabular}{@{}lrrrrrrr@{}}\toprule
\textbf{Tool} & \textbf{\emph{Contamination}} & \textbf{\emph{SARS-CoV-2}} & \textbf{\emph{E. coli}} & \textbf{\emph{Yeast}} & \textbf{\emph{Green Algae}} & \textbf{\emph{Human}} & \textbf{\emph{Relative Abundance}} \\\midrule
\multicolumn{8}{c}{CPU Time (sec)} \\\midrule
\unc         & 265,902.26                     & 36,667.26 & 35,821.14 & 8,933.52 & 16,769.09 & 262,597.83 & 586,561.54 \\
\sig         & 4,573.18                       & 1,997.84 & 23,894.70 & 11,168.96 & 31,544.55 & 4,837,058.90 & 11,027,652.91 \\
\mech        & 3,721.62                       & 1,832.56 & 8,212.17 & 4,906.70 & 25,215.23 & 2,022,521.48 & 4,738,961.77 \\\midrule
\multicolumn{8}{c}{Real time (sec)} \\\midrule
\unc & 20,628.57 & 2,794.76 & 1,544.68 & 285.42 & 2,138.91 & 8,794.30 & 19,409.71 \\
\sig & 6,725.26 & 3,222.32 & 2,067.02 & 1,167.08 & 2,398.83 & 158,904.69 & 361,443.88 \\
\mech & 3,917.49 & 1,949.53 & 957.13 & 215.68 & 1,804.96 & 65,411.43 & 152,280.26 \\\midrule
\multicolumn{8}{c}{Peak memory (GB)} \\\midrule
\unc & 0.65 & 0.19 & 0.52 & 0.37 & 0.81 & 9.46 & 9.10 \\
\sig & 111.69 & 28.26 & 111.11 & 14.65 & 29.18 & 311.89 & 489.89 \\
\mech & 4.13 & 4.20 & 4.16 & 4.37 & 11.75 & 52.21 & 55.31 \\\bottomrule
\end{tabular}}
\label{tab:mapping_resources}
\end{table*}

\subsection{Mapping Time per Read}\label{subsec:mappingtimeperread}

\begin{figure}[tbh]
  \centering
  \includegraphics[width=0.6\columnwidth]{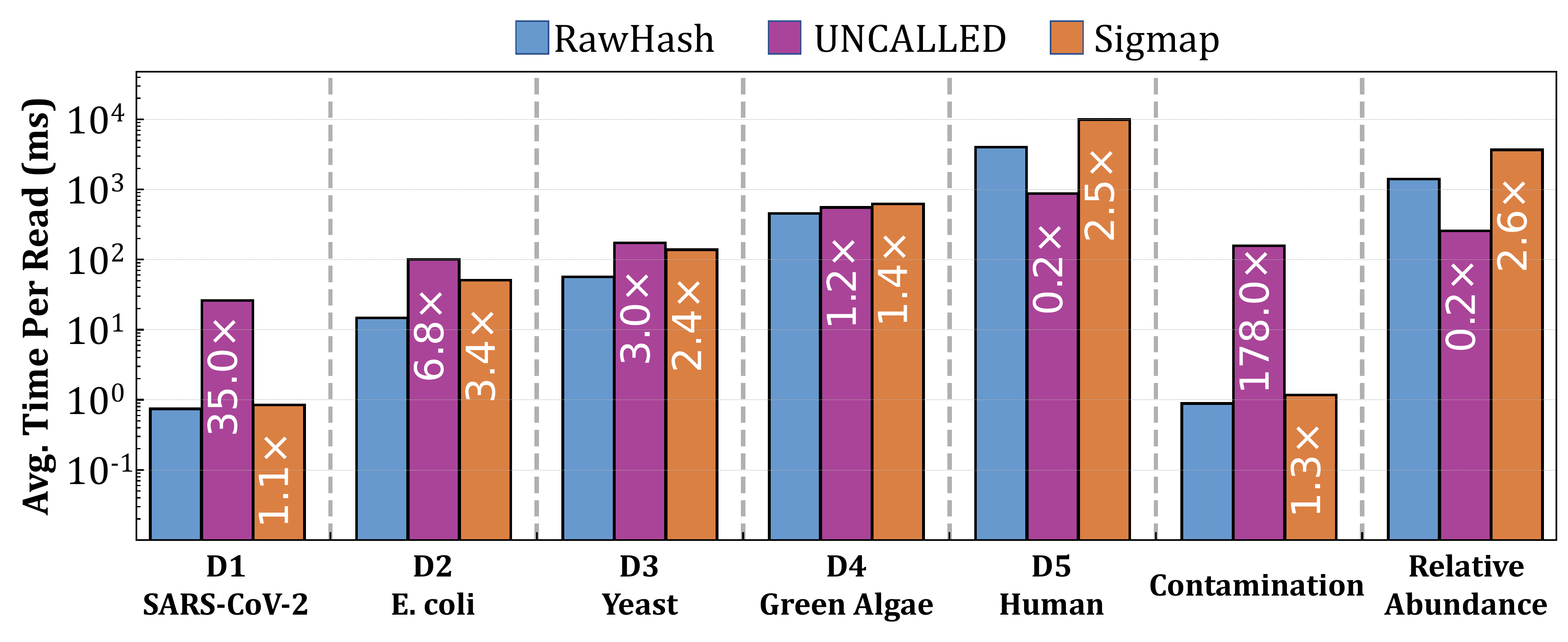}
  \caption{Average time spent per read by each tool in real-time. Values inside the bars show the speedups that \mech provides over other tools in each dataset.}
  \label{fig:timeperread}
\end{figure}

\rev{Supplementary Figure~\ref{fig:timeperread} shows the average mapping time that each tool spends per read for all the datasets we evaluate.}

\section{Limitations of \mech} \label{sec:limitations}

We find four limitations of \mech, which we believe can be improved with further optimizations and better solutions.
First, \mech depends on previously generated k-mer models to generate events from reference genomes. Although these k-mer models can be trained and generated~\citesupp{supp_simpson_detecting_2017, supp_nanopolish_kmers}, this makes it challenging to adapt the most accurate parameters for each k-mer model based on the nanopore model used for sequencing. A more generic k-mer model that can accurately represent all nanopores is needed to easily adapt \mech to all possible nanopore models that may be released in the future.

Second, \mech starts providing lower recall values as the genome size increases, which indicates that a larger portion of correct reads cannot be mapped by \mech due to the increase in the number of false negatives. Although such an increase in false negatives does not substantially affect some applications, such as contamination analysis, where providing higher precision is more critical to correctly identify the contaminated sample, improving it is useful to provide more accurate genome analysis overall.

Third, we perform our relative abundance estimations based on a priori knowledge of reference genomes. While such an experiment can still be useful in practical scenarios, this is not the common case in metagenomic analysis, where a sample is searched against a significantly larger set of species. We expect that our mechanism can still scale to such metagenomic analyses given that many metagenomic databases are efficiently constructed by including fewer and useful information for each species~\citesupp{supp_breitwieser_krakenuniq_2018}, as opposed to our analysis, where we include whole-genome references. 

Fourth, we observe that the throughput of \mech is expected to reach the throughput of a nanopore when analyzing reference genomes slightly larger than a human genome. Such a limitation can be alleviated by applying 1)~seeding techniques that provide faster and more space-efficient searches in large spaces and 2)~chaining algorithms that are optimized for hash-based seed matches without the notion of distance between seeds, unlike the chaining algorithm used in Sigmap.

\clearpage
\section{Configuration} \label{sec:configuration}
\subsection{Parameters} \label{subsec:parameters}

\Copy{R1/2B}{\rev{In Supplementary Table~\ref{tab:parameters}, we show the parameters of each tool for each dataset. In Supplementary Table~\ref{tab:presets}, we show the details of the preset values that \mech sets in Supplementary Table~\ref{tab:parameters}. For \unc, \sig, and minimap2, we use the same parameter setting for all datasets. For the sake of simplicity, we only show the parameters that we explicitly set in each tool. For the descriptions of all the other parameters, we refer to the help message that each tool generates, including \mech.

We note that the parameter names shown in Supplementary Table~\ref{tab:versions} are different from the parameters explained in Sections~\ref{subsec:quantization} and \ref{subsec:hashvalues}, although these parameters essentially perform in the same way as explained in these sections, which we describe next. First, the quantization parameter, $Q$ in Section~\ref{subsec:quantization}, is set using the \texttt{-q} parameter. Second, the value of $p$ in Section~\ref{subsec:quantization} (i.e., pruned bits) can be calculated as $p = Q -l -3$ where $l$ is the least significant $l$ bits of $Q$. We use $l$ instead of $q$ due to its easier programmability in our tool. This $l$ value is set using the \texttt{-l} parameter in \mech. Third, the number of events packed together, $n$ in Section~\ref{subsec:hashvalues}, is set using the \texttt{-e} parameter.

We set these \texttt{-q}, \texttt{-l}, and \texttt{-e} parameters empirically for three types of datasets: 1) viral genomes, 2) small genomes (i.e., $< 50M$ bases, and 3) large genomes (i.e., $> 50M$ bases) using the preset values \texttt{-x viral}, \texttt{-x sensitive}, and \texttt{-x fast}, respectively. In our empirical analysis, we identify that accuracy and performance are significantly impacted by the values we set for \texttt{-e}, \texttt{-q} and \texttt{-l} for three reasons.
First, \texttt{e} determines the number of quantized event values packed in a single hash value. Packing a larger number of events improves the sensitivity as it becomes more challenging to find larger consecutive matches of quantized event values than finding a smaller number of consecutive matches. Finding a smaller set of matches can decrease the time spent in seeding and chaining, as we explain in Supplementary Section~\ref{sec:profiling}.
Second, these values determine the level of quantization of actual event values. Smaller \texttt{-q} and \texttt{-l} values can lead to loss of information due to storing only fewer bits that cannot be useful for identifying significantly different events. Larger \texttt{-q} and \texttt{-l} values can generate different quantized values for highly similar event values that may be corresponding to the same DNA content.
Third, the number of bits that we store for each event, which are determined by \texttt{-q} and \texttt{-l}, impacts the number of events that can be packed in a single 32-bit or 64-bit value. Packing a larger number of events in a single hash value directly impacts sensitivity as discussed earlier in this paragraph (first point).}}

\begin{table*}[tbh]
\centering
\caption{\Copy{R1/2C}{\rev{Parameters we use in our evaluation for each tool and dataset in mapping.}}}
\resizebox{\linewidth}{!}{\begin{tabular}{@{}lccccccc@{}}\toprule
\textbf{Tool} & \textbf{\emph{Contamination}} & \textbf{\emph{SARS-CoV-2}} & \textbf{\emph{E. coli}} & \textbf{\emph{Yeast}} & \textbf{\emph{Green Algae}} & \textbf{\emph{Human}} & \textbf{\emph{Relative Abundance}} \\\midrule
\mech         	 & -x viral -t 32             & -x viral -t 32             & -x sensitive -t 32      & -x sensitive -t 32    & -x fast -t 32               & -x fast -t 32         & -x fast -t 32\\\midrule
\unc  			 & \multicolumn{7}{c}{map -t 32} \\\midrule
\sig        	 & \multicolumn{7}{c}{-m -t 32} \\\midrule
Minimap2         & \multicolumn{7}{c}{-x map-ont -t 32} \\\bottomrule
\end{tabular}
}
\label{tab:parameters}
\end{table*}

\begin{table*}[tbh]
\centering
\caption{\Copy{R1/2D}{\rev{Corresponding parameters of presets (-x) in RawHash.}}}
\resizebox{0.6\linewidth}{!}{\begin{tabular}{@{}lcc@{}}\toprule
\textbf{Preset (-x)} & \textbf{Corresponding parameters} & Usage \\\midrule
viral         	 & -e 5 -q 9 -l 3 & Viral genomes \\\midrule
sensitive  	     & -e 6 -q 9 -l 3 & Small genomes (i.e., $<50M$ bases)\\\midrule
fast        	 & -e 7 -q 9 -l 3 & Large genomes (i.e., $>50M$ bases)\\\bottomrule
\end{tabular}
}
\label{tab:presets}
\end{table*}

\subsection{Versions}\label{subsec:versions}

\Copy{R1/3A}{\rev{Supplementary Table~\ref{tab:versions} shows the version and the link to these corresponding versions of each tool that we use in our experiments.}}

\begin{table*}[tbh]
\centering
\caption{\Copy{R1/3B}{\rev{Versions of each tool.}}}
\resizebox{\linewidth}{!}{\begin{tabular}{@{}lll@{}}\toprule
\textbf{Tool} & \textbf{Version} & \textbf{Link to the Source Code} \\\midrule
\mech & 0.9 & \url{https://github.com/CMU-SAFARI/RawHash/tree/8042b1728e352a28fcc79c2efd80c8b631fe7bac}\\\midrule
\unc  & 2.2 & \url{https://github.com/skovaka/UNCALLED/tree/74a5d4e5b5d02fb31d6e88926e8a0896dc3475cb}\\\midrule
\sig  & 0.1 & \url{https://github.com/haowenz/sigmap/tree/c9a40483264c9514587a36555b5af48d3f054f6f}\\\midrule
Minimap2 & 2.24 & \url{https://github.com/lh3/minimap2/releases/tag/v2.24}\\\bottomrule
\end{tabular}
}
\label{tab:versions}
\end{table*}

\clearpage

% \end{document}

\let\noopsort\undefined
\let\printfirst\undefined
\let\singleletter\undefined
\let\switchargs\undefined

\bibliographystylesupp{IEEEtran}
\bibliographysupp{main}
\end{document}